### K. Lodders
Department of Earth and Planetary Sciences,
Washington University, St. Louis, Missouri 63130, USA

### H. Palme
Sektion Meteoritenforschung
Forschungsinstitut und Naturmuseum Senckenberg
Senckenberganlage 25, D-60325 Frankfurt am Main, Germany

### H.-P. Gail
Center for Astronomy, Institute of Theoretical Astrophysics,
University of Heidelberg, Albert-Überle-Str. 2, 69120 Heidelberg, Germany

# Chapter 3.4 Abundances of the elements in the solar system

## 3.4.1 Introduction

### 3.4.1.1 Historic remarks

The first serious attempt to establish a comprehensive list of "cosmic abundances" was made by the geochemist Victor Moritz Goldschmidt. During the 20s and 30s of the last century Goldschmidt and his colleagues in Göttingen, and later in Oslo, measured and compiled chemical data of terrestrial rocks, meteorites and individual phases of meteorites. Based on these data Goldschmidt [38G] set up a cosmic abundance table which he published in 1938 in the ninth volume of his "Geochemische Verteilungsgesetze der Elemente" (The Geochemical Laws of the Distribution of the Elements) entitled "Die Mengenverhältnisse der Elemente und der Atom-Arten" (The proportions of the elements and the various types of atoms). Like Farrington in 1915, Goldschmidt believed that meteorites might provide the average composition of cosmic matter. He realized that most meteorites should be representative of average cosmic matter because they have not been affected by physico-chemical processes (e.g., melting and crystallization), whereas the crust of the Earth, which formed after formation of a metal core by melting of the mantle, is not representative of the Earth's bulk composition. The bulk Earth composition should be that of average cosmic matter. Goldschmidt calculated the average concentrations of elements in cosmic matter by using a weighted mean of element abundances in meteorite phases: metal (2 parts), sulfide (1 part) and silicates (10) parts. Since the highly volatile elements (C, O, N, rare gases), are depleted in meteorites, Goldschmidt amended his abundance list with Russell's abundance data of the solar photosphere that had become available in 1929 [29R]. In his list of solar abundances, Russell incorporated earlier work on meteorites by Goldschmidt to derive the abundance distribution of the elements.

Goldschmidt's and Russell's lists had most of the major features of the solar system abundances as we know them today: The dominance of H and He, the strong decrease in abundance with increasing atomic number, the very low abundances of Li, Be, and B, and the pronounced abundance peak at $^{56}$Fe. These lists also confirmed the significantly higher abundances of elements with even atomic number than that of their odd-numbered neighbors which had been discovered before by Oddo and Harkins in the 1910s and 20s for elements up to the Fe-peak. In his paper, Goldschmidt [38G] commented briefly on the existence of transuranian elements. He also recognized the preponderance of nuclides with mass numbers of 50 and 82 which among other things led to the nuclear shell model.

Almost 20 years later Brown [49B] and Suess and Urey [56S] published new abundance tables. The Suess and Urey compilation was particularly influential for theories of nucleosynthesis ([57B, 57C]) and for the development of nuclear astrophysics in general. Later compilations ([73C, 89A, 93P, 03P1, 03L] and others) took into account improved analytical data of meteorites and of the solar photospheric spectrum.

The German physicist Joseph von Fraunhofer (1787-1826) was the first to observe dark lines in the spectrum of the light emitted from the Sun and he identified several hundred lines. Some fifty years later G. Kirchhoff (1824-1887) recognized that the dark lines in the sunlight spectrum are characteristic of chemical elements and he correctly concluded that the outer cooler layers of the Sun absorb radiation from the hotter underlying parts. Sodium was the first element that was detected in terrestrial rocks, meteorites, and the Sun. At the beginning of the 20th century, Miss C.H. Payne made the first attempts to quantify the information contained in stellar and solar absorption lines. She recognized the large abundance of H in stars and gave the first estimates of the chemical composition of stellar atmospheres. Later N. H. Russell [29R] published the first list of the elemental abundances of 56 elements in the solar photosphere. Improvements in instrumentation and in interpretation of the absorption spectra through increased knowledge of atomic absorption properties as well as better modeling of the solar photosphere produced increasingly accurate data of the composition of the solar photosphere.

The photospheric abundances of elements heavier than oxygen (and except the rare gases) can be compared with the abundances in meteorites. As discussed below, abundances of five meteorites constituting the group of CI-chondrites show excellent agreement with photospheric abundances. This agreement has evolved over the years and is now better than ±10% for 39 elements as described below. The combination of elemental abundances in CI-meteorites and in the Sun allows an estimate of the elemental composition of the bulk solar system.

### 3.4.1.2. Solar system matter

Our solar system is the result of the gravitational collapse of a small part of a giant molecular cloud. It is often assumed that the Sun, the planets, and all other objects in the solar system formed from a hot gaseous nebula with well defined chemical and isotopic composition. The findings of comparatively large (on a per mil level) and widespread variations in oxygen isotopic composition has cast doubt upon this assumption (see [03C] and references therein). Additional evidence of incomplete mixing and small-scale homogenization of elements and nuclides in the primordial solar nebula is provided by the detection of huge (up to the percent level) isotope anomalies of some heavy elements in meteoritic silicon carbide (SiC) grains (e. g., [03Z, 05L1]).

Until recently it was still assumed that isotope anomalies delivered by pre-solar grains are volumetrically negligible and that in all types of meteorites and planets, elements other than oxygen are isotopically uniform, with the exception of some rare Ca, Al-rich inclusions (CAI) that have large isotopic anomalies in most elements analysed (FUN-inclusions). Rotaru et al. [92R] first showed that although Cr in bulk carbonaceous chondrites is isotopically uniform, selective dissolution of meteorites revealed large isotopic variations in the various fractions thus removed from the meteorites. Similar observations were later made for Os [05B1] and Zr [05S1].

As an example, the isotopic composition of Zr in various fractions of the Murchison meteorite is shown in Fig. 1a [05S1]. The fractions were produced by dissolving the meteorite with increasingly stronger acids. The different symbols indicate the leaching steps (top grey square is thefirst, black symbol at bottom is the last leaching step; white circle is bulk meteorite). When compared to the normal terrestrial isotopic composition of Zr, the first leachates have an excess of r-process $^{96}$Zr, the last ones show a depletion. The latter pattern is very similar to that of Zr in some graphite and mainstream SiC-grains that originated in red giant stars, whereas the first patterns are comparable to patterns found in graphite grains and SiC grains of type X believed to come from supernovae (Figs. 1b-d). The anomalies in the separated presolar grains are orders of magnitude higher than in the meteoritic leachates; notice the different size of the scales in Fig. 1. Although individual phases in chondritic meteorites carry anomalies in Zr (and some other elements) bulk meteorite samples have the same isotopic composition as other solar system materials, including the Earth. In a few cases, such as, for example $^{54}$Cr, anomalies may show up in bulk samples. So far it appears that these anomalies are restricted to elements that form stable compounds at high temperatures or are incorporated into such phases. The important conclusion is that the solar nebula never was entirely isotopically homogeneous, which also excludes that all components of chondrites condensed from a single homogeneous gaseous cloud.

Primitive chondrites may contain a significant fraction of unprocessed interstellar material with large isotope anomalies in physically separable components. However, there must be an intimate mixture of components on a very fine scale, as the anomalies show only small effects (Cr) or do not show up at all in bulk samples (Os, Zr). The larger variations in oxygen isotopes may result from extensive gas solid reaction in the early stages of the evolution of the solar system. Chemical similarities of estimated bulk compositions and almost identical isotopic compositions of Earth, Mars, Moon and meteorites indicate a single source for all these objects, the proto-solar nebula, well mixed on a macroscopic scale. Still, there are subtle chemical variations in the bulk composition of chondritic meteorites as well as planets (e.g. volatile elements) that may be ascribed to fractionation during formation and processing of solid solar system matter, as described in the next section.

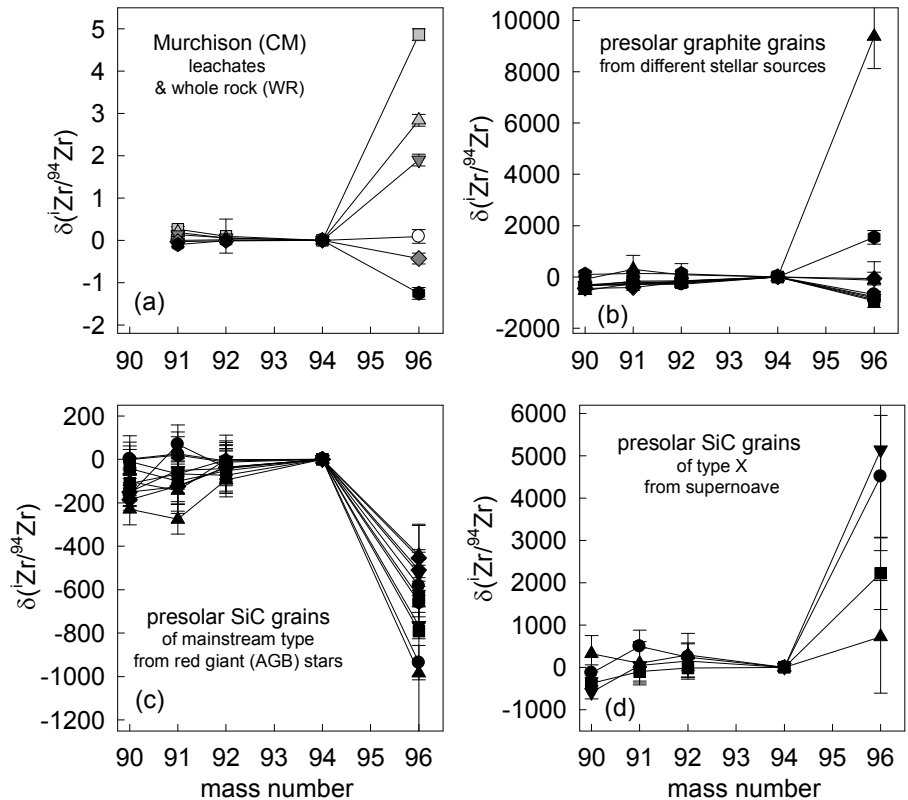

**Fig. 1.** (a) Isotopic variations of Zr in fractions of the Murchison meteorite produced by successive leaching with increasingly stronger acids, see text [05S1]. The isotope $^{96}$Zr, mainly made by rapid neutron capture (r-process), is variable, which indicates different Zr isotopic compositions in different phases of the meteorite. These phases cannot have formed from an isotopically uniform solar nebula. Bulk samples of meteorites usually do not show an anomaly. Grains with excesses and depletions of $^{96}$Zr must balance on a sub-micrometer scale. (b-d) Qualitatively similar but much stronger anomalies are observed in presolar grains recovered from meteorites. Both, SiC and graphite grains with excess $^{96}$Zr (r-process) and depletion of $^{96}$Zr are present. Presolar grain data from [97N, 98N, 06P]. $\varepsilon_{Zr}$ is the deviation of the observed $^{96}$Zr/$^{90}$Zr-ratio from the solar system standard ratio in 1/10000.

### 3.4.1.3 Classification of solar system materials

Meteorites recovered from the surface of the Earth only spent a few million years as individual bodies in space as determined from the abundance of certain cosmic-ray-produced isotopes, mostly rare gases. Before that, meteorites were buried in the interiors of larger planetesimals. Two different types of such planetesimals may be distinguished, those which were once molten and differentiated into an iron core and a silicate mantle and those which were not heated sufficiently to reach melting temperatures. The latter are called chondritic meteorites because most of them contain mm-sized spherules, the chondrules, which were molten in space before accretion into the meteorite parent body. Chondritic meteorites preserved the bulk composition of their parent planetesimal and their comparatively uniform composition should, to a first order, reflect the average composition of the solar system, except for extremely volatile elements.

A variety of meteorite types is derived from differentiated planetesimals, iron-meteorites from the core (or segregated metal-pods), stony-irons from the core-mantle boundary and basaltic meteorites from the crust of their parent planetesimal. These meteorites are, obviously, less useful in deriving average solar system abundances.

Chondritic meteorites appear to be derived from a comparatively uniform source, but there are chemical variations among them that indicate processing in the solar nebula prior to accretion, such as incomplete condensation, evaporation, preferred accumulation and separation of metal by magnetic forces, differential movement of fine vs. coarse grained material, etc. On the meteorite parent bodies, aqueous alteration and mobilisation of volatiles at metamorphic temperatures may have contributed to variations in bulk composition, although to a lesser extent.

### 3.4.1.4 Condensation temperatures

Many of the processes responsible for the variable chemical composition of primitive meteorites are related to the temperature of formation of meteoritic components. Although it is difficult to unambiguously establish the condensation origin of any single meteoritic component, the relative elemental abundance patterns in meteorites indicate that condensation and/or evaporation processes must have occurred in the early solar system and that the nebular volatilities of the elements were an important factor in establishing the various meteorite compositions. A convenient measure of volatility is the condensation temperature. These temperatures are calculated by assuming thermodynamic equilibrium between solids and a cooling gas of solar composition. Major elements condense as minerals while minor and trace elements condense in solid solution with the major phases. The temperature where 50% of an element is in the condensed phase is called the 50% condensation temperature [85W]. Condensation temperatures depend on total pressure. Table 1 lists the calculated 50% condensation temperatures at a nebular total pressure of $10^{-4}$ bars taken from [03L].

Within this framework, six more or less well defined components may account for the variations in the elemental abundances in primitive meteorites. In addition, the state of oxidation of a meteorite is an important parameter that controls abundance variations, which adds to the complexity of chondritic meteorites.

### 3.4.2. Chondritic meteorites

### 3.4.2.1. Components of chondritic meteorites

According to the condensation temperatures, the following components are distinguished (Table 1):

(1) *Refractory component:* The first phases to condense from a cooling gas of solar composition are Ca, Al-oxides and silicates associated with a comparatively large number of trace elements, such as the REE (rare earth elements), Zr, Hf, and Sc (Table 1). These elements are often named refractory lithophile elements (RLE), in contrast to the refractory siderophile elements (RSE) comprised of metals with low vapor pressures, e.g., W, Os, Ir (Table 1) that condense at similarly high temperatures as multi-component metal alloys. Both RLE and RSE are enriched in CAI by a factor of twenty on average, indicating that the refractory component makes up about 5 % of the total condensable matter. *Variations in e.g., Al/Si, Ca/Si ratios of bulk chondrites may be ascribed to the incorporation of variable amounts of an early condensed refractory phase.*

(2) *Mg-silicates:* The major fraction of condensable matter is associated with the three most abundant elements heavier than oxygen: Si, Mg and Fe (see below). In the relatively reducing environment of the solar nebula, Fe condenses almost entirely as metal alloy (FeNi), while Mg and Si form forsterite ($Mg_2SiO_4$) which converts to enstatite ($MgSiO_3$) at lower temperatures by reaction with gaseous SiO. The overall atomic abundances of Mg and Si are about equal, and since the forsterite stoichiometry gives an atomic Mg/Si ratio twice the solar system ratio, forsterite is an ideal candidate for producing variations in Mg/Si-ratios among chondrites. *Variations in Mg/Si ratios of bulk meteorites are produced by the incorporation of various amounts of early-formed forsterite.*

**Table 1.** Cosmo- and geochemical character of the elements and 50% condensation temperatures [K]

| | Refractory elements | | | | | Moderately volatile elements | | | |
|---|---|---|---|---|---|---|---|---|---|
| | litho. | sidero. | chalco. | host phase | | litho. | sidero. | chalco. | host phase |
| Re | | 1821 | | rfm-alloy | Pd | | 1324 | | FeNi-alloy |
| Os | | 1812 | | rfm-alloy | Li | 1142 | | | forst.+enst. |
| W | | 1789 | | rfm-alloy | As | | 1065 | | Fe-alloy |
| Zr | 1764 | | | $ZrO_2$ | Au | | 1060 | | Fe-alloy |
| Hf | 1703 | | | $HfO_2$ | Cu | | 1037 | | Fe-alloy |
| Sc | 1659 | | | hibonite | Ag | | 996 | | Fe-alloy |
| Y | 1659 | | | hibonite | Sb | | 979 | | Fe-alloy |
| Gd | 1659 | | | hibonite | Ga | | 968 | | FeNi+feld. |
| Tb | 1659 | | | hibonite | Na | 958 | | | feldspar |
| Dy | 1659 | | | hibonite | B | 908 | | | feldspar |
| Ho | 1659 | | | hibonite | Ge | | 883 | | Fe-alloy |
| Er | 1659 | | | hibonite | Rb | 800 | | | feldspar |
| Tm | 1659 | | | hibonite | Te | | 709 | | Fe-alloy |
| Lu | 1659 | | | hibonite | Cs* | 799 | | | feldspar |
| Th | 1659 | | | hibonite | Bi* | | 746 | | Fe-alloy |
| Al | 1653 | | | hibonite | F | 734 | | | F-apatite |
| U | 1610 | | | hibonite | Pb* | | 727 | | Fe-alloy |
| Ir | | 1603 | | rfm-alloy | Zn | 726 | | | forst.+enst. |
| Nd | 1602 | | | hibonite | Te | 709 | | | FeNi-alloy |
| Mo | | 1590 | | rfm-alloy | Sn | 704 | | | FeNi-alloy |
| Sm | 1590 | | | hib + tit | Se | | | 697 | troilite |
| Ti | 1582 | | | titanite | S | | | 664 | troilite |
| Pr | 1582 | | | hib + tit | | | | | |
| La | 1578 | | | hib + tit | | Highly volatile elements | | | |
| Ta | 1573 | | | hib + tit | | | | | |
| Nb | 1559 | | | titanite | | litho. | sidero. | chalco. | |
| Ru | | 1551 | | rfm-alloy | Cd | 652 | | | enst.+troil. |
| Ca | 1659 | | | hib + gehl | Br | 546 | | | Cl-apatite |
| Yb | 1517 | | | hib + tit | In | | | 536 | FeS |
| Ce | 1487 | | | hib + tit | I | 535 | | | Cl-apatite |
| Sr | 1478 | | | titanite | Tl | | | 532 | troilite |
| Be | 1464 | | | melilite | Hg | | | 252 | troilite |
| V | 1452 | | | titanite | | | | | |
| Pt | | 1408 | | rfm-alloy | | Ices and gases | | | |
| Rh | | 1392 | | rfm-alloy | O | | 180 | | rock + ice |
| Eu | 1356 | | | hib+tit+fel | N | | 123 | | $NH_3·H_2O$ |
| | Elements in major components | | | | Xe | | 68 | | $Xe·6H_2O$ |
| | litho. | sidero. | chalco. | host phase | Kr | | 52 | | $Kr·6H_2O$ |
| Si | | 1354 | | forsterite | C | | 40 | | $CH_4·7H_2O$ + $CH_4$ ice |
| Mg | | 1354 | | forsterite | | | | | |
| Fe | | 1334 | | FeNi-alloy | Ne | | 9.1 | | Ne ice |
| Co | | 1352 | | FeNi-alloy | H | | 182 | – | – |
| Ni | | 1353 | | FeNi-alloy | He | | < 3 | – | – |

Calculated 50% condensation temperatures, assuming indicated host phase(s). * - behaviour in meteorites suggests highly volatile element (see text for details)

(3) *Metallic iron condenses* as a FeNi-alloy at about the same temperature as forsterite, the sequence depending on total pressure. At pressures above $10^{-4}$ bar, Fe-metal condenses before forsterite and at lower pressures forsterite condenses ahead of metal. *Variations in the concentrations of Fe and other siderophile elements in meteorites are produced by the incorporation of variable fractions of metal.*

(4) *Moderately volatile elements* have condensation temperatures between those of Mg-silicates and FeS (troilite). The most abundant of the moderately volatile elements is sulfur which starts to condenses by reaction of gaseous $H_2S$ with solid Fe at 710 K, independent of total pressure. Half of all sulfur is condensed by 664 K Other moderately volatile elements condense in solid solution with major phases. Moderately volatile elements are distributed among sulfides, silicates and metal. *Relative depletions of moderately volatile elements in meteorites are caused by incomplete condensation. The amount and the relative abundances of these elements in meteorites are probably the result of removal of volatiles during condensation [88P].*

(5) *Highly volatile elements* have condensation temperatures below that of FeS and above water ice (Table 1). The group of *highly volatile elements* comprises elements with very different geochemical affinities, such as the chalcophile Pb and the lithophile I. *Similar processes as those invoked for the depletion of moderately volatile elements are responsible for variations in these elements. In addition, heating on small parent bodies may lead to loss of highly volatile elements.*

*(6) Ultra volatile elements* have condensation temperatures below that of water ice. This group includes H, C N, O, and the noble gases. About 20 % of the total O is condensed at high temperatures when oxides and silicates are stable but the abundances of rock-forming cations are too small to sequester larger amounts of O. Only condensation of water ice can lead to more than 50% condensation of O.
In Fig. 2, the depletion of moderately volatile elements in CM-chondrites relative CI-chondrites is shown. The smooth decline in abundance ratios with condensation temperature - irrespective of the geochemical character of the elements - suggests volatility control of abundances in CM-chondrites.

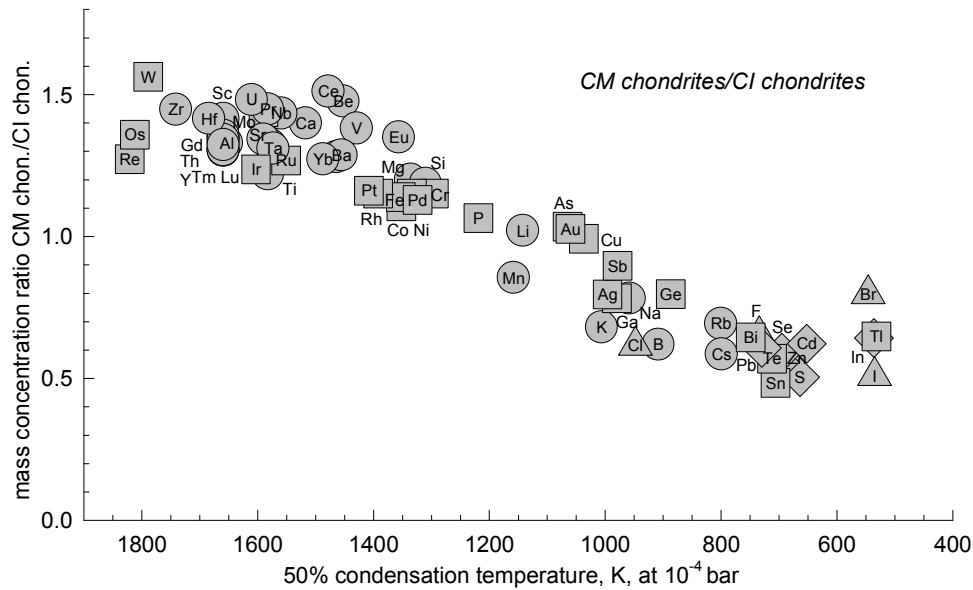

**Fig. 2.** Mass concentration ratios of the elements in CM-chondrites over CI-chondrites as a function of condensation temperature. CI-chondrites have the highest concentrations of moderately volatile and highly volatile elements. All other groups of meteorites are lower. The smooth decline in abundance ratios with condensation temperature irrespective of the geochemical character of the elements suggests volatility controlled abundances in CM-chondrites. The symbol shape indicates the geochemical character of the element: square = siderophile, circle = lithophile, diamond = chalcophile, triangle = halogen.

**3.4.2.2 Chemical variations in chondritic meteorites and the significance of CI-chondrites**

The variations of selected element/silicon concentration ratios in the different groups of chondritic meteorites are shown in Fig. 3. All element/silicon ratios are further normalized to the CI-ratios (taken from Table 3). Meteorite groups are arranged in order of decreasing bulk oxygen content, i.e. decreasing average oxygen fugacity (Fig. 3). Element ratios in the solar photosphere determined by absorption spectroscopy are shown for comparison (Table 4).

The element Al is generally representative of refractory elements and the Al/Si ratios indicate the fraction of the refractory component relative to the major element fraction in the meteorite groups. The Al/Si ratio of CI-chondrites agrees best with the photospheric ratio, although the ratios in CM (type 2) carbonaceous chondrites and even OC (ordinary chondrites) are almost within the error bar of the solar ratio, which is above 20%. The errors of the meteorite ratios are below 10 %, in many cases below 5 %. A very similar pattern across the meteorite groups in Fig. 3 as for Al/Si would be obtained for other refractory element/Si ratios (Ca, Ti, Sc, REE etc.), because the ratios among refractory elements are more or less constant in all classes of chondritic meteorites. The average Sun/CI-meteorite ratio from 20 well determined refractory lithophile elements (Al, Ca, Ti, V, Sr, Y, Zr, Nb, Ba, La, Ce, Pr, Nd, Sm, Eu, Gd, Dy, Er, Lu, see Table 4) is 1.007 with a standard deviation of 5.4%. Thus the photospheric/CI-chondritic ratio of refractory elements has an uncertainty of only 5.4%. It appears that only CI-chondrites match the refractory element/Si ratios of the Sun. The relative concentration of refractory elements is higher in all types of carbonaceous chondrites (except CR-chondrites, not shown in Fig. 3) and lower in ordinary chondrites and enstatite chondrites (EC).

The Mg/Si ratio of CI-chondrites matches that of the Sun (Fig. 3). This ratio is, however, less diagnostic, because carbonaceous chondrite groups have almost the same Mg/Si ratio [01W]. The OC and the EC contain significantly less Mg. The error of the solar ratio is about 20% (combining the errors of Si and Mg) and covers the range of all classes of chondrites except the EH-chondrites.

The range of Fe in chondritic meteorites varies by about a factor of two (Fig. 3). Recently discovered sub-groups of carbonaceous chondrites show large excesses or depletions of Fe, indicating that metal behaves as an independent component; some groups are depleted in Fe and other siderophile elements, others are enriched, independent of other properties (e.g. [03K]). The excellent agreement of CI-meteorite abundance ratios with solar abundance ratios is obvious, although the formal error of the solar Fe-abundance is significant (Table 4). However, here the same argument applied to refractory elements can be used. The abundance patterns of Ni and Co are identical to those of Fe and the error in the solar abundances of Ni is only half of that of Fe and Co (Table 4). The combined error associated with the solar/CI-meteorite ratio for these three metals is then below 10 %, demonstrating that most types of meteorites have metal/silicate ratios different from CI-meteorites.

The CI-chondrites have the highest contents of moderately volatile and highly volatile elements (Na, Zn, S) and are highest in oxygen of all chondritic meteorites (a large fraction of O is in hydrous silicates and hydrated salts). However, in contrast to other highly volatile elements, such as Pb, the content of oxygen is still a factor of two below that of the solar photosphere (Fig. 3), implying that water is not fully condensed. The notion of incomplete condensation of volatile elements in most meteorite groups is supported by the observation that there is no known group of meteorites enriched in these elements relative to the average solar system abundances [88P].

Only one group of meteorites, the CI-chondrites closely matches solar abundances for elements representing the various cosmochemical groups, except for the extremely volatile elements such as the rare gases, H, C, N, O, and Li which will be discussed in a later section. All other chondrite groups deviate from solar abundances and these deviations can be understood, at least in principle, by gas-solid fractionation processes in the early solar system.

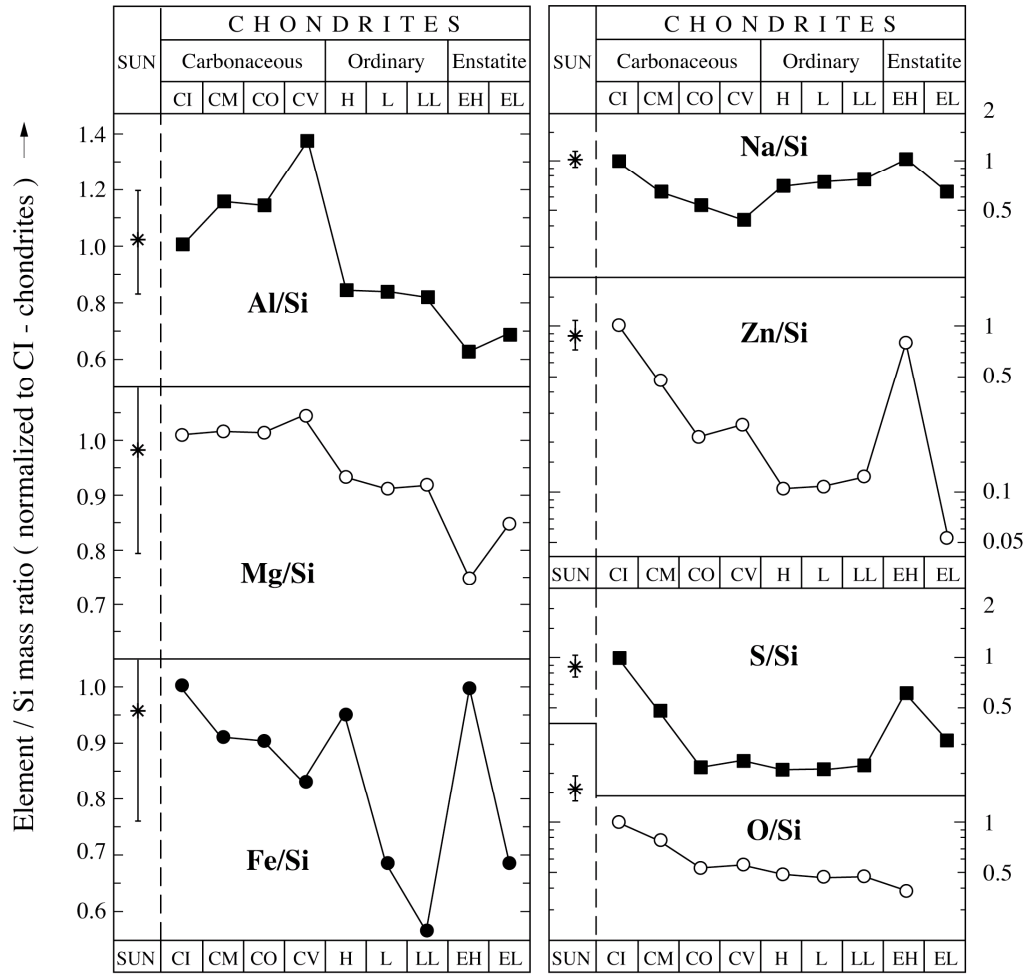

**Fig. 3**. Element/Si mass ratios of characteristic elements in various groups of chondritic (undifferentiated) meteorites. Chondrite groups are arranged according to decreasing oxygen content. The refractory component is represented by Al; Mg represents the Mg-silicates, Fe the metal component and Na, Zn and S the moderately volatile elements. Only CI-chondrites fit with photospheric abundances indicated on the left side. See text for details. Data from [88W].

### 3.4.3 Meteorite derived solar system abundances

#### 3.4.3.1 CI-meteorites as a standard for solar system abundances

Among the more than forty thousand recovered meteorites, only five are observed falls of CI-chondrites: Orgueil, Ivuna, Alais, Tonk, and Revelstoke (Table 2). These meteorites are very fragile and easily break up during atmospheric entry. In addition, their survival time on Earth is short, which is why CI-chondrites found in Antarctica are of limited use as abundance standards. Most analyses have been performed on the Orgueil meteorite, simply because it is the largest CI-meteorite and material is easily available for analysis. However, problems with sample size, sample preparation and the mobility of the elements are reflected in the chemical heterogeneities within the Orgueil meteorite, and can make a comparison of the data obtained by different authors difficult. This contributes to the uncertainties in CI abundances.

**Table 2.** Observed CI-chondrite falls

| Meteorite | Date of fall | Country | Preserved mass |
|---|---|---|---|
| Alais | 15 March 1806 | France | 6 kg |
| Ivuna | 16 Dec. 1938 | Tanzania | 0.7 kg |
| Orgueil | 14 May 1868 | France | 14 kg |
| Revelstoke | 31 March 1965 | Canada | <1 g |
| Tonk | 22 Jan. 1911 | India | 10 g |

The chemical composition of CI-chondrites as shown Fig. 3 is the basis for designating these meteorites as primitive or unfractionated. Texturally and mineralogically, they are far from being primitive. CI-meteorites are microbreccias with millimeter to submillimeter clasts of variable composition. Late stage fractures filled with carbonates, hydrated Ca and Mg sulfates demonstrate that low temperature processes have affected the meteorite, and are even occurring presently, in the terrestrial environment, such as redistribution of some elements into epsomite (hydrated Mg-sulfate). The CI-chondrites have no chondrules and consist almost entirely of extremely fine grained hydrous silicates with about 10% magnetite, ~5-7% salts (carbonates, sulfates), some pyrrhotite and pentlandite (less than 5%; [04B1, 06M1]). High temperature phases such as olivine and pyroxene make only up a few percent.

Clearly, the CI-chondrites match solar abundances very closely, their present texture and mineralogy has been largely established by processes that occurred late on the Orgueil parent body. However, on a cm scale these processes must have been essentially isochemical, otherwise the chemical composition of Orgueil would be much more heterogeneous and would not have retained solar abundance ratios for so many elements.

**3.4.3.2 Abundances of the elements in CI-chondrites**

The abundance Table is an update of an earlier compilation in Landolt-Börnstein by [93P] and [03P]. These compilation emphasise the use of the Orgueil meteorite as CI standard rock, because it is the most massive of the 5 CI-chondrite falls and therefore the most analyzed one. Another approach was taken by [03L] who used data from all 5 CI-chondrites and computed weighted average compositions. However, since most analytical data are from Orgueil, these weighted averages are strongly dominated by this meteorite. In the compilation here we continue the tradition of [93P] and [03P] using primarily Orgueil data for CI-abundances. A comparison of the data from Orgueil and other CI-chondrites is in [03L] and a more detailed comparison of analytical data of CI-chondrites is made in a forthcoming paper. However, compositional differences among CI-chondrites are very small, if they exist at all (see Table 3 in [03L]).

During the last years, many new data on trace elements became available, primarily because of improvement in instrumentation. The application of Inductively Coupled Plasma Mass Spectrometry (ICP-MS) has led to a wealth of new, high quality analyses of Orgueil and other primitive meteorites. These data are included, if the new results are sufficiently different from the earlier abundances (Table 3). In many cases an extensive collection of data on CI-chondrites by Lodders is used [09L].

One of the problems with analytical data on CI-chondrites is the variable absolute abundance level of many elements, due to variable water contents and/or massive alteration of fine grained matrix material. It is therefore sometimes useful to consider ratios of geochemically similar elements and tie them to the absolute abundance of a single element, which needs to be accurately determined. For example, the Ca/Al ratio of 1.085 is fairly constant in carbonaceous chondrites [70A]. In the new compilation the average Al content of 25 Orgueil analyses is 0.850 % with a standard deviation of 6.4%, whereas the average Ca content of 0.901 % has a standard deviation of 14%, twice that of Al [09L]. We therefore used the **Al** content of 0.850 % and multiplied it with the carbonaceous chondrite Ca/Al ratio to obtain 0.922 % of **Ca** for average Orgueil. The reason for the larger spread in Ca is most likely the higher mobility of Ca in the aqueous environment of the Orgueil parent body.

Another example is the Y/Ho ratio that has been determined in carbonaceous chondrites as 25.94 with an uncertainty of 0.3 % [07P], but the absolute concentrations of Y and Ho are much more uncertain. We used such ratios to infer more precise concentrations, in other instances we checked the consistency of the Orgueil concentrations derived from average values by using constraints from element ratios.

Each element was assigned an estimated uncertainty. In cases where the quality of individual analyses is roughly comparable, this reflects the spread in individual Orgueil analyses. In cases with limited analyses we use the uncertainty given by the authors. The distinction between accuracy and precision is not clear in all cases. Some authors prefer to give only estimates for precision. The listed uncertainties of Table 3 should give a rough estimate of the quality of the CI-values, but for elements where precision is high but accuracy low we have assigned larger uncertainties than those given by authors. Here concentrations for Orgueil are given in ppm and ppb by mass, which is also abbreviated as μg/g and ng/g, respectively. For most elements we list three significant digits. Some of the data were taken from earlier compilations: [93P], [03P1], [03L], and for several elements marked "com" in Table 3, we added new literature data or made new estimates as described in the text.

The light element abundances (Li to N) are taken from [03L] and we included additional data from [06M2] for **Be**. Oxygen in meteorites is difficult to determine and especially complicated in CI-chondrites because of the hygroscopic nature of the Mg-sulfates that form as alteration products in the terrestrial environment. The varying water contents - depending on humidity - and increased oxidation of original meteoritic phases change the **O** content of CI-chondrites while they reside in museums. Three direct analysis of the O concentrations by fast neutron activation analysis gave 46.54±0.8 mass% for Orgueil ([09S1], [64W]), whereas the O content computed as the difference to 100% gives 45.85 mass%. The 1.5 % higher INAA value may well reflect the varying degrees of hydration. Here the oxygen value of 45.85 mass% from the difference to 100% is listed in the Table, but O variations due to changing water content should be borne in mind when using this value for any modelling.

There are many analyses of **Na** in Orgueil. By removing those which deviate by more than 10 % from the average a mean Na of 4990 ppm is obtained. The average **Mg** and **Si** contents from 20 Orgueil analyses are 9.58 % and 10.70 % respectively, leading to a typical Si/Mg carbonaceous chondrite ratio of 0.895, half a percent below 0.900. For Al see discussion in the introduction to this section.

The average **P** abundance in Orgueil from 9 analyses is 967 ± 90 ppm. The average is mainly based on XRF data [01W] and [90J] and recent ICPMS data by [06M2].

For **Ti** we obtain a value of 451±37 ppm, based on an average of 17 analyses [09L]. For elements from V to Ge newer results are similar to the data in [93P] which we have kept, except for **Fe**, which is now 18.53 % [09L].

We adopted 1.74 ppm for **As** from an average of 20 analyses [09L]. We primarily used INAA data for Se, which average to 20.3 ppm [09L]. Preference to INAA was given for the Br concentration, however, following [82A] data from [79D] were not included.

We used primarily mass spectrometric data for **Rb** and **Sr**, which give 2.31 ppm Rb and 7.81 ppm Sr, using 21 and 20 analyses [09L], respectively.

Eleven analyses give an average value of 1.58±0.15 for Y [09L]. In a second approach we considered the very constant Y/Ho ratio of 25.94±0.84 [07P]. A similar value of 25.9 is derived from data in [06M2]. Using the **Ho** concentration of 0.0752 ppm (see below) gives an Y concentration of 1.48 ppm. We use the mean of both values and obtain 1.53 ppm for Y.

The average **Zr** of 18 Orgueil samples is 3.82 ppm with a spread of 0.42 ppm [09L]. This is nearly identical to the old value of 3.86 ppm by [00J]. The ICP-MS Orgueil analysis of 4.935 ppm by [03M] is much too high, probably reflecting sample heterogeneity. Other ICP-MS analyses of four small Orgueil chips yielded an average concentration of 3.69 ± 0.42 ppm [07L1], again with a significant spread that suggests inhomogeneous distribution of Zr in Orgueil. The average Zr/Hf of 14 carbonaceous chondrites analyzed by [03M] is 34.09 with a standard deviation of 1.3%. The paper by [07L1] reports a similar ratio of 33.97 ± 0.9 for the average from four Orgueil chips. In addition [79S] determined a mean Zr/Hf ratio of 34.3 ± 0.2 in CM-meteorites and reported the same ratio for CV-meteorites. In view of this consistency, we calculate a Zr concentration of 3.62 ppm from the average Zr/Hf= 34.12 using the relatively well defined Hf content of 106 ppb (see below). The low Zr content for Orgueil needs confirmation.

The Zr/Nb ratio of 13.08 ± 1.06 measured by [03M] is reasonably constant in all carbonaceous chondrites and agrees with the average ratio of 12.78 ± 0.25 in four Orgueil samples analyzed by [07L1]. An average Zr/Nb ratio of 12.93 yields a **Nb** concentration of 0.279 ppm for CI-chondrites. This compares well to the direct average of 0.280±0.037 ppm from 11 analysis but both values are significantly higher than the earlier value of 0.247 ppm based on [00J]. These authors had obtained a 20 % higher Zr/Nb ratio. We prefer the newer ICP-MS analyses of [03M] and [07L1] and list a Nb concentration of 0.279 ppm.

Analyses for **Mo** are given by [89B], [81P], [00W], and [07L1]. The average Mo content of 0.944 ppm in 4 small Orgueil samples analyzed by [07L1] is close to earlier estimates of Mo in CI-chondrites. The average from all these sources gives 0.973 ± 0.107 ppm Mo in Orgueil [09L].

The mean **Ru** concentration of 18 Orgueil samples is 686 ± 44 ppb, which leads to a Ru/Os ratio of 1.39 (see below for Os). The mean carbonaceous chondrite ratio by [03H] is 1.40 ± 0.06 in agreement with the average Orgueil data compiled by [09L]. For Rh there is little choice than to use 139 ppb from [95J] and [96J]. The estimated uncertainty for **Rh** was arbitrarily increased to account for accuracy.

The **Pd** data of 17 Orgueil samples average to 558 ± 19 ppb. Several analyses by thermal ionisation mass spectrometry (TIMS) average to 555 ppb [84L], in agreement with 556 ppb determined with laser ablation (LA) -ICP-MS by [08S1]. Two ICP-MS analyses by [96Y] give 570 ppb and two determinations by [03H] yield 574 ppb.

New average concentrations were calculated for elements from Ag to Te. Any differences to earlier values were found to be very small in all cases. Iodine in Orgueil from 4 samples by [79D] and one from Goles et al. 1967 [67G] averages to 530+-120 ppb. The Cs and Ba values have not changed significantly, so we kept the old values from [93P].

Early REE analyses of Orgueil were done with neutron activation procedures, whereas modern analyses use mass spectrometric methods. The latter are generally considered more accurate. We have primarily used data sets by [74N], [84B], and [06M2]. Two Orgueil analyses with isotope dilution (ID) by [78E] are untypical for Orgueil, one is too high in all REE analysed, the other sample is too low. However, average Orgueil calculated from these and other chondrite data by [78E] fit reasonably well with the average calculated from the data sets mentioned above. Some of the REE analyses by [02F] deviate considerably from the averages in [09L], in particular Dy, Yb and Eu (see below). The [02F] data were therefore not considered in the averages. We used REE concentration ratios of CI- and CM-chondrites for some of the REE ([74N], [06M2]), which improves statistics. The inclusion of CM-chondrite data makes the explicit assumption that the refractory REE are not fractionated from each other in CM- chondrites. The values of all REE concentrations derived by these procedures agree with the older values given in [03P] within 2 %. Estimated uncertainties of all REE are set to 5 % for multi-isotope elements and 7 % for mono-isotopic REE.

We used **Yb** as reference element in calculating other REE abundances from element ratios because the Orgueil compilation of [09L] gives a mean Yb of 169 ppb with a comparatively small spread of 5 % for 22 analyses. Two groups analysed REE with isotope dilution (ID) and thermal ionisation mass spectrometry (TIMS), [74N] and found Yb concentrations of 169 ppb and [84B] 170 ppb, respectively. New Yb results with ICP-MS give 175 ppb [06M2]. This is slightly above the old neutron activation values by the UCLA and Mainz groups [93P]. Six mass-spectrometric and two INAA data sets average to an Yb content of 169 ppb, which we use in further discussions. Based on the spread of six individual analyses we estimate an uncertainty of about 5%, consistent with the standard deviation from all 22 values.

The **La** concentration of 242 ppb in Orgueil is the average from 19 La analyses by different methods [09L]. This value is in perfect agreement with the La concentration derived from the Yb/La ratio of 0.699 obtained from 6 Orgueil ([74N], [84B], [06M2]) and 4 Murchison ([74N], [06M2]) samples. Similar procedures were applied to **Ce**, **Sm**, **Gd**, **Dy**, and **Er**. The averages from [09L] and the values obtained from Yb/element correlations in CI- and CM-chondrites are used to produce the concentrations listed in Table 3. The difference between the two data sets is below 2% in most cases. This procedure gives most weight to data obtained by mass spectrometry but considers also the data produced with other methods. The Sm/Nd ratio of 0.323 derived from the data sets above and the extensive set of [08B1] leads to a Nd concentration of 471 ppb. The concentrations of mono-isotopic REE were not determined by [74N] and [84B], and their determination with INAA is also more difficult. The ICP-MS analyses of [06M2] produced data on **Pr**, **Tb**, **Ho** and **Tm**. The CI-abundances of these elements were established by taking the mean of the concentrations derived by the method using Yb/element concentration ratios with the data by [06M2] and for CM-chondrites, and the concentration averages from [09L].

The average for **Lu** from 21 analyses gives 24.92 ± 1.7 ppb in Orgueil, whereas the mean of 6 TIMS and ICP-MS analyses is 25.48 ± 0.0006. Using the Yb/Lu ratio of 6.67 derived from ID and ICP-MS data, gives a Lu content of 25.34 ± 1.3 ppb for Orgueil. The mean of the three approaches is 25.3 ppb, marginally higher than the average from all data, which include INAA values as well.

The Lu/Hf ratio of 0.238 is extremely well determined in chondritic meteorites because dating with the $^{176}$Lu-$^{176}$Hf system requires knowledge of precise concentrations of both elements. Applying our recommended Lu concentration of 25.3 ppb, this ratio translates to an **Hf** concentration of 106 ppb. The average of 24 individual analyses by INAA and mass spectrometry for Orgueil gives 110±10 ppb Hf. Giving preference to mass spectrometric methods also leads to an Hf concentration of 106 ppb for CI-chondrites. The old isotope dilution analysis of Hf by [84B] gave 106.6 ppb as an average of two samples. A recent analysis by ICP-MS [04K] showed some variability in Hf with one sample containing 106.4 ppb in good agreement with the result of another analysis with 106 ppb by [08B1]. A second sample from [08B1], however, contained 120.9 ppb, which is not included in averaging. In Table 3 we list a value of 106 ppb with an estimated uncertainty of 5 %.

As mentioned above, two recent data sets for Zr, Nb, Hf and **Ta** yield the same Zr/Hf and Zr/Nb ratios, but they disagree in their Nb/Ta ratios. While [03M] report Nb/Ta = 20.40 in CI- and CM-chondrites, [07L1] obtained an average of 17.99 from their four Orgueil analyses. This is close to the average Nb/Ta ratio of 17.73 by [00J] for three CM- and CI-chondrites. We adopt the average Nb/Ta ratio of 19.20 from [03M] and [07L1] and calculate the corresponding Ta content of 0.0145 ppm for CI-chondrites, which is close to the old Ta values of 0.0142 in [03P] and 0.0145 ppm in [03L], but less than the average of 0.0159+-0.0018 ppm which results by including all new data. The high standard deviation of the average shows that for Ta the ratio method is more reliable. The abundances of Zr, Hf, Nb and Ta are not yet fully consistent. A particular problem is that [03M] found a significantly lower Nb/Ta ratio in type 3 carbonaceous chondrites than in CI- or CM-chondrites, implying fractionations among refractory elements within the various groups of carbonaceous chondrites. This is not corroborated by other refractory element ratios, such as the Zr/Hf and Zr/Nb ratios which are constant in the various groups of carbonaceous chondrites.

The **W** concentration in CI-chondrites can be calculated from Hf/W ratios. Several recent analyses are available for carbonaceous chondrites because the Hf/W ratio is important for radiometric dating. Both Hf and W are refractory elements and their concentration ratio appears to be constant within the various types of carbonaceous chondrites. The average Hf/W ratio of two Orgueil samples, five Allende splits and 12 other carbonaceous chondrites analysed by [04K] is 1.11±0.13 ppm. A Hf content of 106 ppb leads to a W CI-abundance of 96 ± 10 ppb, in agreement with earlier data by [82R], and the direct average of 96±13 ppb from 7 analyses [09L].

The noble metals include Au, Ag, and the six Platinum Group Elements (PGE) Ru, Rh, Pd and Os, Ir, Pt. Except for Pd, the PGE are refractory metals with generally constant concentration ratios in chondritic meteorites. The refractory noble metals and Re condense at temperatures above FeNi-alloys, whereas the noble metals Pd, Au and Ag have condensation temperatures below FeNi [08P], [03L]. The major reason for variable Pd/Ir and Au/Ir ratios in different types of chondrites is the higher volatility of Pd and Au. In geochemistry, the noble metals and Re are collectively termed highly siderophile elements (HSE), because their metal/silicate partition coefficients are above 10,000.

| | | This work | | | | Anders & Grevesse 1989 [89A] | % difference |
|---|---|---|---|---|---|---|---|
| | Element | Mean CI (ppm by mass) | Estimated accuracy (%) | Source | Atoms per $10^6$ Si atoms | Mean CI (ppm by mass) | [89A] - this work |
| 1 | H | 19700 | 10 | [03L] | 5.13E+06 | 20200 | 2.4 |
| 2 | He | 0.00917 | | [03L] | 0.601 | 0.00917 | 0.0 |
| 3 | Li | 1.47 | 13 | [03L] | 55.6 | 1.50 | 2.0 |
| 4 | Be | 0.0210 | 7 | com | 0.612 | 0.0249 | 18.6 |
| 5 | B | 0.775 | 10 | com | 18.8 | 0.870 | 12.3 |
| 6 | C | 34800 | 10 | [03L] | 7.60E+05 | 34500 | -0.9 |
| 7 | N | 2950 | 15 | [03L] | 55300 | 3180 | 7.9 |
| 8 | O | 459000 | 10 | com | 7.63E+06 | 464000 | 1.2 |
| 9 | F | 58.2 | 15 | [03J] | 804 | 60.7 | 4.3 |
| 10 | Ne | 0.00018 | | [03L] | 0.00235 | 0.000180 | 0.0 |
| 11 | Na | 4990 | 5 | com | 57000 | 5000 | 0.2 |
| 12 | Mg | 95800 | 3 | com | 1.03E+06 | 98900 | 3.3 |
| 13 | Al | 8500 | 3 | com | 82700 | 8680 | 2.1 |
| 14 | Si | 107000 | 3 | com | 1.00E+06 | 106400 | -0.6 |
| 15 | P | 967 | 10 | com | 8190 | 1220 | 26.2 |
| 16 | S | 53500 | 5 | [03L] | 4.38E+05 | 62500 | 15.5 |
| 17 | Cl | 698 | 15 | [93P] | 5170 | 704 | 0.9 |
| 18 | Ar | 0.00133 | | [03L] | 0.00962 | 0.00133 | 0.0 |
| 19 | K | 544 | 5 | [03P] | 3650 | 558 | 2.6 |
| 20 | Ca | 9220 | 5 | com | 60400 | 9280 | 0.7 |
| 21 | Sc | 5.9 | 5 | [93P] | 34.4 | 5.82 | -1.4 |
| 22 | Ti | 451 | 8 | com | 2470 | 436 | -3.3 |
| 23 | V | 54.3 | 5 | [93P] | 280 | 56.5 | 4.1 |
| 24 | Cr | 2650 | 3 | [93P] | 13400 | 2660 | 0.5 |
| 25 | Mn | 1930 | 3 | [93P] | 9220 | 1990 | 2.9 |
| 26 | Fe | 185000 | 3 | com | 8.70E+05 | 190400 | 2.8 |
| 27 | Co | 506 | 3 | [93P] | 2250 | 502 | -0.8 |
| 28 | Ni | 10800 | 3 | [93P] | 48300 | 11000 | -6.2 |
| 29 | Cu | 131 | 10 | [93P] | 541 | 126 | -3.8 |
| 30 | Zn | 323 | 10 | [93P] | 1300 | 312 | -3.4 |
| 31 | Ga | 9.71 | 5 | [93P] | 36.6 | 10.0 | 3.0 |
| 32 | Ge | 32.6 | 10 | [93P] | 118 | 32.7 | 0.3 |
| 33 | As | 1.74 | 9 | com | 6.10 | 1.86 | 6.9 |
| 34 | Se | 20.3 | 7 | com | 67.5 | 18.6 | -8.4 |
| 35 | Br | 3.26 | 15 | com | 10.7 | 3.57 | 9.5 |
| 36 | Kr | 5.22E-05 | | [03L] | 1.64E-04 | 5.22E-05 | 0.0 |
| 37 | Rb | 2.31 | 7 | com | 7.10 | 2.30 | -0.4 |
| 38 | Sr | 7.81 | 7 | com | 23.4 | 7.80 | -0.1 |
| 39 | Y | 1.53 | 10 | com | 4.52 | 1.56 | 2.0 |
| 40 | Zr | 3.62 | 10 | com | 10.4 | 3.94 | 9.1 |

Table 3. Abundances in CI-chondrites based on the Orgueil meteorite

| | | Table 3. Abundances in CI-chondrites based on the Orgueil meteorite | | | | |
|---|---|---|---|---|---|---|
| | | This work | | | Anders & Grevesse 1989 [89A] | % difference |
| Element | | Mean CI (ppm by mass) | Estimated accuracy (%) | Source | Atoms per $10^6$ Si atoms | Mean CI (ppm by mass) | [89A] - this work |
| 41 | Nb | 0.279 | 10 | com | 0.788 | 0.246 | -11.8 |
| 42 | Mo | 0.973 | 10 | com | 2.66 | 0.928 | -4.6 |
| 44 | Ru | 0.686 | 6 | com | 1.78 | 0.712 | 3.8 |
| 45 | Rh | 0.139 | 10 | com | 0.355 | 0.134 | -3.6 |
| 46 | Pd | 0.558 | 5 | com | 1.38 | 0.560 | 0.4 |
| 47 | Ag | 0.201 | 5 | com | 0.489 | 0.199 | -1.0 |
| 48 | Cd | 0.674 | 7 | com | 1.57 | 0.686 | 1.8 |
| 49 | In | 0.0778 | 7 | com | 0.178 | 0.080 | 2.8 |
| 50 | Sn | 1.63 | 15 | com | 3.60 | 1.720 | 5.5 |
| 51 | Sb | 0.145 | 15 | com | 0.313 | 0.142 | -2.1 |
| 52 | Te | 2.28 | 7 | com | 4.69 | 2.320 | 1.8 |
| 53 | I | 0.530 | 20 | com | 1.10 | 0.433 | -18.3 |
| 54 | Xe | 1.74E-04 | | com | 3.48E-04 | 1.74E-04 | 0.0 |
| 55 | Cs | 0.188 | 5 | [93P] | 0.371 | 0.187 | -0.5 |
| 56 | Ba | 2.41 | 6 | [93P] | 4.61 | 2.340 | -2.9 |
| 57 | La | 0.242 | 5 | com | 0.457 | 0.2347 | -3.0 |
| 58 | Ce | 0.622 | 5 | com | 1.17 | 0.6032 | -3.0 |
| 59 | Pr | 0.0946 | 7 | com | 0.176 | 0.0891 | -5.8 |
| 60 | Nd | 0.471 | 5 | com | 0.857 | 0.4524 | -3.9 |
| 62 | Sm | 0.152 | 5 | com | 0.265 | 0.1471 | -3.2 |
| 63 | Eu | 0.0578 | 5 | com | 0.100 | 0.0560 | -3.1 |
| 64 | Gd | 0.205 | 5 | com | 0.342 | 0.1966 | -4.1 |
| 65 | Tb | 0.0384 | 7 | com | 0.0634 | 0.0363 | -5.5 |
| 66 | Dy | 0.255 | 5 | com | 0.412 | 0.2427 | -4.8 |
| 67 | Ho | 0.0572 | 7 | com | 0.0910 | 0.0556 | -2.8 |
| 68 | Er | 0.163 | 5 | com | 0.256 | 0.1589 | -2.5 |
| 69 | Tm | 0.0261 | 7 | com | 0.0406 | 0.0242 | -7.3 |
| 70 | Yb | 0.169 | 5 | com | 0.256 | 0.1625 | -3.8 |
| 71 | Lu | 0.0253 | 5 | com | 0.0380 | 0.0243 | -3.6 |
| 72 | Hf | 0.106 | 5 | com | 0.156 | 0.104 | -1.9 |
| 73 | Ta | 0.0145 | 10 | com | 0.0210 | 0.0142 | -2.1 |
| 74 | W | 0.096 | 10 | com | 0.137 | 0.0926 | -3.5 |
| 75 | Re | 0.0393 | 10 | com | 0.0554 | 0.0365 | -7.1 |
| 76 | Os | 0.493 | 8 | com | 0.680 | 0.486 | -1.4 |
| 77 | Ir | 0.469 | 5 | com | 0.640 | 0.481 | 2.6 |
| 78 | Pt | 0.947 | 8 | com | 1.27 | 0.990 | 4.5 |
| 79 | Au | 0.146 | 10 | com | 0.195 | 0.140 | -4.1 |
| 80 | Hg | 0.350 | 20 | [09L] | 0.458 | 0.258 | -26.3 |
| 81 | Tl | 0.142 | 8 | [09L] | 0.182 | 0.142 | 0.0 |
| 82 | Pb | 2.63 | 7 | [09L] | 3.33 | 2.470 | -6.1 |
| 83 | Bi | 0.110 | 9 | [09L] | 0.138 | 0.114 | 3.6 |
| 90 | Th | 0.0310 | 8 | [09L] | 0.0351 | 0.0294 | -5.2 |
| 92 | U | 0.00810 | 8 | [09L] | 0.00893 | 0.0081 | 0.0 |

The average of 28 **Re** analyses of individual Orgueil samples is 39.3 ± 3.7 ppm and 30 samples give 493 ppm **Os** with a spread of ± 38 ppm [09L]. The resulting Os/Re ratio is 12.54, which is in excellent agreement with 12.27 ± 0.43 from the Os/Re ratio of 17 carbonaceous chondrite samples analysed by [03H]. The average **Ir** content of 44 Orgueil samples is 469 ± 25 ppb. The larger number of values reflects that Ir is the only PGE that can be reliably determined by instrumental neutron activation analysis at the level present in chondritic meteorites. The Os/Ir ratio of 1.053 compares favourably with the average of 1.05 ±0.02 from type 1 and 2 carbonaceous chondrites [03H]. The mean **Pt** concentration of 17 Orgueil samples is 947 ± 72 ppb. The corresponding Pt/Ir ratio is 2.02, which is slightly above the ratio of 1.93 ± 0.1 obtained from the CI- and CM-chondrites [03H]. The average concentration of 146 ±17 ppb for **Au** is from 40 analyses of Orgueil [09L].

### 3.4.3.3 Comparison with other abundance tables

The chemical composition of the Orgueil meteorite is now known better than 10% for most elements. There are 8 elements where the difference between the element concentrations in Orgueil recommended here and those of the Anders and Grevesse compilation [89A] differ by more than 10% (see Table 3 for a comparison). New analyses since 1989 have improved the accuracy of the concentrations of Be, B, P, S, Se and Nb, while the abundances of I and Hg are still poorly known. Orgueil is heavily contaminated with Hg [93P] and for I new bulk analyses are necessary to resolve the existing differences. There are 20 elements which differ by less than 10% but by more than 5%. These are all trace elements where, in most cases, new analyses have produced more consistent results (Br, Zr, Nd, Tm, Re, As, Ni, Pb, Pr, Sn, Tb, and Th). For the three mono-isotopic REE Tm, Pr, and Tb there are new data with ICP-MS, although these were done on small samples. As discussed above, the concentration values of these three elements were further improved by using concentration ratios with other elements in CM-meteorites. The Zr content is lower than in most previous compilations, and the new value is based on the very constant Zr/Hf ratio of carbonaceous chondrites [03M]. However, the Zr concentration needs further study. The Re content in Orgueil is now better defined from the constant Os/Re ratio measured in different carbonaceous chondrites and the reasonably well defined Os-content, which in turn is based on a constant Os/Ir ratio. The Br and N concentrations in Orgueil and the other CI-meteorites may be quite variable.

## 3.4.4 Photospheric abundances

Most elemental abundance data for the Sun are derived by spectroscopy of the solar photosphere. The quantitative determination of the elemental abundances in the sun involves three steps, first the construction of a numerical model atmosphere, then the calculation of an emitted spectrum based on the model atmosphere, finally a comparison of this spectrum with the observed spectrum, e.g., [95C]. Converting the absorption lines into abundances requires the knowledge of line positions of neutral and ionized atoms and the transition probabilities and lifetimes of the excited atomic states. Another assumption usually made in calculating solar abundances is that of local thermodynamic equilibrium (LTE), i.e. "the quantum-mechanical states of atoms, ions, and molecules are populated according to the relations of Boltzmann and Saha, valid strictly in thermodynamic equilibrium" [01H]. In newer calculations deviations from local thermodynamic equilibrium, ("NLTE") are taken into account. More recent hydrodynamical models include 3D effects such as convective flows and granulation of the solar atmosphere. The application of 3D models instead of 1D or 2D models to abundance determinations led to lower abundances of several elements, notably oxygen, but significant reductions of photospheric abundances in other elements (e.g., Na, Al, Si) were also reported (see, e.g., [05A1]). In addition, different 3D models lead to different results, as discussed, for example, by Shi et al. [08S2] for silicon. Hence abundances derived with 3D models should be regarded with some caution. The lower abundances derived for some elements from these models cause problems for standard solar models that describe the evolution of the sun to its current radius and luminosity.

In the following we give preference to elemental abundances derived with more conservative solar atmospheric models; although for some elements (e.g., P, S, Eu) 1D/2D and some 3D models produce the same results (see below).

Our recommended solar photospheric abundances are listed in Table 4. Elemental abundances are normalized to $10^{12}$ atoms of hydrogen, usually on a logarithmic scale as practice in astronomy. The data are given as:

$$A(El) = \log N(El)/N(H) + 12$$

where N(El)/N(H) is the atomic ratio of element El to H. As mentioned above, most elements are determined spectroscopically. Exceptions are rare gases and some elements that have no accessible or only heavily blended lines for quantitative spectroscopy in the photosphere (e.g., As, Se, Br, Te, I, Cs). For some elements, sunspot spectra (e.g., F, Cl, Tl, In) are used and the uncertainty of these data is large. Most of the data listed in Table 4 are from an earlier compilation by Lodders [03L] where references to the original papers can be found. Uncertainties are given in the logarithmic scale as dex-units, which are connected to the standard deviation σ by:

$$\sigma \text{ (in \%)} = (10^{\sigma \text{ (in dex)}} - 1) \times 100$$

Elements for which new abundances are available are discussed separately in the next section.

**Li:** We kept the value of A(Li) = 1.1±0.1 selected in [03L] which was determined by [94C1]. In [05A1] the analysis from Mueller et al. [75M] of A(Li)=1.0±0.1 was amended with 3D models to give 1.05±0.10. Considering the large uncertainty involved and the uncertainties in the 3D models, we keep the [03L] value.

**Be:** A detailed discussion of the difficulties associated with the determination of the photospheric and meteoritic Be abundance is given in [03L]. A subsequent re-analysis by Asplund [04A1] gave A (Be) = 1.38±0.09 for the photosphere, suggesting basic agreement between photospheric and meteoritic Be abundances. Apparently Be destruction has not occurred over the Sun's lifetime.

**C**: The C abundance from Allende Prieto et al. [02A] was selected in [03L]. This value was confirmed by Asplund et al. [05A2] and by a recent analysis of CO by Scott et al. [06S].

**N**: The N abundance is still uncertain. We take the results of a detailed study by Caffau et al. [09C].

**O:** The recommended O abundance is an average from the recent determinations by [08C1], [08L2], and [08M1]. A detailed review on the problem of the solar O abundance and other light elements is given by Basu and Anita [08B2]. Adopting A(C) = 8.39 and A(O) = 8.73±0.07 gives a C/O ratio of C/O = 0.457, less than the C/O ratio of 0.50 in more recent studies and compilations, but still somewhat higher than the value of C/O = 0.427 from [89A].

**Ne**: Results by Morel and Butler [08M2] from Ne I and Ne II lines in nearby, early B-type stars yield A(Ne) = 7.97±0.07, which can be taken as characteristic of the present day ISM. If neon contributions from more massive AGB stars to the ISM are negligible, this value may be taken as representative of the Sun 4.6 Ga ago. Landi et al. [07L2] determined A(Ne) = 8.11±0.1 from solar flare measurements in the ultra-violet. The average from these studies gives A(Ne) = 8.05±0.06, but considering the larger uncertainties in the determinations, an overall uncertainty of 0.10 dex (25%) is easily warranted for the recommended value here.

Bahcall et al. [05B2] concluded that a neon abundance of A(Ne) = 8.29±0.05 could restore the agreement of standard solar models and helioseismological observations that existed before the year 2000, when more sophisticated photospheric modelling began to yield lower C, O and other heavier element abundances. However, using Ne to restore opacity lost by C and O requires increasing the Ne abundance by a factor of ~2.6 (~0.42 dex) from the Ne values in [03L] or [05A1], which is quite high. If we adopt A(Ne) = 8.05 from above, and consider that the O abundance is also slightly higher than previously estimated, solar models may be more consistent with helioseismological constraints.

**Na:** The selected Na value of 6.30±0.03 is from the [03L] compilation. [05A1] found A(Na) = 6.17±0.04 from six Na lines and their 3D model atmospheres. We do not adopt this value, as it is ~25% lower than the meteoritic value as well as previously determined photospheric Na abundances.

**Al:** [05A1] give A(Al) = 6.37±0.06, which is lower than the corresponding meteoritic value. The previous value of 6.47±0.07 from 1D models selected in [03L] is much better in agreement with meteoritic data. The Al/Si from 3D models in [05A1] is 1.2 times that of CI-chondrites.

**Si**: The value selected here is from Shi et al. [08S2].

**P:** The value of A(P)=5.49±0.04 in [03L] is changed to the recent result of A(P) = 5.46±0.04 by Caffau et al. [07C1]. The new value is based on 3D atmospheric models, but Caffau et al. find that the determination of P by 1D models is not significantly different. The new value agrees well with the meteoritic value.

**S:** New determinations with 3D modeling by Caffau [07C1] and [07C2] give 7.14±0.01. Results from 1D models are very similar.

**Ar**: The value of A(Ar) is derived from various independent sources since the Ar abundance cannot be determined spectroscopically in the photosphere [08L1].

**K:** The K abundance used in [03L] was confirmed by [06Z]. The value proposed by [05A1] is too low, a similar situation as for Na.

**Ca**: The recommended value is from the LTE analysis by Reddy et al. [03R]. The value from 3D models in [05A1] is about 5% lower. A new Ca analysis in the photosphere using several different models is needed.

**Sc**: The Sc abundance remains uncertain and Zhang et al. [08Z] recommend a range of $3.07 \leq A(Sc) \leq 3.13$. We take A(Sc) = 3.10 with an appropriate uncertainty of 0.1 dex, which covers the range of values reported in recent papers.

**Cr:** Sobeck et al. [07S] determined A(Cr) = 5.64±0.01, which is identical with the photospheric value listed in [03L] but with much lower uncertainty.

**Mn:** Two recent studies give A(Mn) = 5.37±0.05 [07B1]) and 5.36±0.10 [07B2]), which are only slightly lower than the value 5.39±0.03 given in [03L]. The LTE analysis by Reddy et al. [03R] of 5.37±0.05 is also in this range. These re-analyses of the photospheric Mn abundance seem to confirm that the photospheric Mn abundance is lower than the meteoritic value of 5.50±0.01.

**Ni**: The updated A(Ni) = 6.23±0.04 from the LTE analysis by Reddy et al. [03R] is similar to the previous value, and has a lower uncertainty.

**Zr:** [06L1] found A(Zr) = 2.58±0.02 from a 3D analysis, 0.01 dex smaller than the previously selected value in [03L].

**Pd**: A new value of A(Pd) = 1.66±0.04 was reported by [06X].

**In:** Gonzales [06G] suggested that the large difference in the In abundances between the photosphere and CI-chondrites may be the result of the relatively high volatility of In. However, there is reasonable agreement between the abundances of other elements of similar volatility in meteorites and the Sun (e.g., S, Tl, Pb). Vitas et al. [08V] obtained an upper limit A(In) = 1.50 and potential blends are considered. A higher than the meteoritic value of A(In) = 0.79 is also unlikely from nucleosynthesis [08V].

**REE** (Rare Earth Elements): Many improvements have been made in the abundance analysis of the REE, in particular through measurements of atomic lifetimes and transition probabilities by the Wisconsin group in a series of detailed papers by Lawler, Sneden, and co-workers [e.g., 03D, 04L1, 06D1, 06L2, 08L3]. A recent paper by Sneden et al. [09S2] summarizes the efforts and gives abundance results for REE. Now the REE abundances are among the best-known abundances for the sun.

**Hf:** The photospheric abundance of Hf A(Hf) = 0.88±0.08 as listed in [03L] was confirmed with a redetermination using improved transition probabilities by Lawler et al. [07L3]. Another recent re-determination of the photospheric Hf abundance using 3D models gives A(Hf) = 0.87±0.04 [08C2].

**Os:** Quinet et al. [06Q] found A(Os) = 1.25±0.11. This is significantly lower than the value of 1.45±0.10 used in previous compilations, including [03L]. Both Os values seem to be problematic when compared to the meteoritic value. The meteoritic/photospheric ratio of the older value is 0.83, and of the new 1.3. Grevesse et al. [07G] selected the newer, smaller abundance. We keep the higher value A(Os) = 1.45 as it fits better to the meteoritic value, and it also fits better with the abundance systematics in the Pt-element region predicted from nucleosynthesis models. A new analysis is required.

**Pt:** According to Den Hartog [05D], the photospheric Pt abundance is not very reliable. The value selected in [03L] is adopted here, but is assigned a 0.3 dex uncertainty (factor 2) to emphasize its low reliability.

**Table 4.** Elemental abundances in the solar photosphere and in CI-chondrites
[log N(H) = A(H) = 12]

|   |    | Solar Photosphere | σ dex | Orgueil CI-chondrite | σ dex | Sun/Meteorite $N_{sun}/N_{met}$ |
|---|----|-------------------|-------|----------------------|-------|---------------------------------|
| 1  | H  | 12     |      | 8.24  | 0.04 | 5710    |
| 2  | He | 10.925 | 0.02 | 1.31  |      | 4.1E+09 |
| 3  | Li | 1.10   | 0.10 | 3.28  | 0.05 | 0.01    |
| 4  | Be | 1.38   | 0.09 | 1.32  | 0.03 | 1.15    |
| 5  | B  | 2.70   | 0.17 | 2.81  | 0.04 | 0.78    |
| 6  | C  | 8.39   | 0.04 | 7.41  | 0.04 | 9.46    |
| 7  | N  | 7.86   | 0.12 | 6.28  | 0.06 | 38.4    |
| 8  | O  | 8.73   | 0.07 | 8.42  | 0.04 | 2.06    |
| 9  | F  | 4.56   | 0.30 | 4.44  | 0.06 | 1.32    |
| 10 | Ne | 8.05   | 0.10 | -1.10 |      | 1.4E+09 |
| 11 | Na | 6.30   | 0.03 | 6.29  | 0.02 | *1.03*  |
| 12 | Mg | 7.54   | 0.06 | 7.55  | 0.01 | *0.98*  |
| 13 | Al | 6.47   | 0.07 | 6.45  | 0.01 | *1.05*  |
| 14 | Si | 7.52   | 0.06 | 7.53  | 0.01 | *0.97*  |
| 15 | P  | 5.46   | 0.04 | 5.45  | 0.04 | *1.03*  |
| 16 | S  | 7.14   | 0.01 | 7.17  | 0.02 | *0.92*  |
| 17 | Cl | 5.50   | 0.30 | 5.25  | 0.06 | 1.79    |
| 18 | Ar | 6.50   | 0.10 | -0.48 |      | 9.6E+06 |
| 19 | K  | 5.12   | 0.03 | 5.10  | 0.02 | *1.06*  |
| 20 | Ca | 6.33   | 0.07 | 6.31  | 0.02 | *1.04*  |
| 21 | Sc | 3.10   | 0.10 | 3.07  | 0.02 | 1.07    |
| 22 | Ti | 4.90   | 0.06 | 4.93  | 0.03 | *0.94*  |
| 23 | V  | 4.00   | 0.02 | 3.98  | 0.02 | *1.05*  |
| 24 | Cr | 5.64   | 0.01 | 5.66  | 0.01 | *0.96*  |
| 25 | Mn | 5.37   | 0.05 | 5.50  | 0.01 | *0.75*  |
| 26 | Fe | 7.45   | 0.08 | 7.47  | 0.01 | *0.95*  |
| 27 | Co | 4.92   | 0.08 | 4.89  | 0.01 | *1.08*  |
| 28 | Ni | 6.23   | 0.04 | 6.22  | 0.01 | *1.03*  |
| 29 | Cu | 4.21   | 0.04 | 4.27  | 0.04 | *0.88*  |
| 30 | Zn | 4.62   | 0.15 | 4.65  | 0.04 | 0.94    |
| 31 | Ga | 2.88   | 0.10 | 3.10  | 0.02 | 0.61    |
| 32 | Ge | 3.58   | 0.05 | 3.60  | 0.04 | *0.95*  |
| 33 | As |        |      | 2.32  | 0.04 |         |
| 34 | Se |        |      | 3.36  | 0.03 |         |
| 35 | Br |        |      | 2.56  | 0.06 |         |
| 36 | Kr | 3.28   | 0.08 | -2.25 |      | 3.4E+05 |
| 37 | Rb | 2.60   | 0.10 | 2.38  | 0.03 | 1.64    |
| 38 | Sr | 2.92   | 0.05 | 2.90  | 0.03 | *1.04*  |
| 39 | Y  | 2.21   | 0.02 | 2.19  | 0.04 | *1.05*  |
| 40 | Zr | 2.58   | 0.02 | 2.55  | 0.04 | *1.07*  |
| 41 | Nb | 1.42   | 0.06 | 1.43  | 0.04 | *0.98*  |

**Table 4.** Elemental abundances in the solar photosphere and in CI-chondrites
[log N(H) = A(H) = 12]

|    |    | Solar Photosphere | σ dex | Orgueil CI-chondrite | σ dex | Sun/Meteorite $N_{sun}/N_{met}$ |
|----|----|---|---|---|---|---|
| 42 | Mo | 1.92 | 0.05 | 1.96 | 0.04 | *0.92* |
| 44 | Ru | 1.84 | 0.07 | 1.78 | 0.03 | *1.14* |
| 45 | Rh | 1.12 | 0.12 | 1.08 | 0.04 | 1.09 |
| 46 | Pd | 1.66 | 0.04 | 1.67 | 0.02 | *0.97* |
| 47 | Ag | 0.94 | 0.30 | 1.22 | 0.02 | 0.52 |
| 48 | Cd | 1.77 | 0.11 | 1.73 | 0.03 | 1.10 |
| 49 | In | <1.50 | UL | 0.78 | 0.03 | 5.21 |
| 50 | Sn | 2.00 | 0.30 | 2.09 | 0.06 | 0.81 |
| 51 | Sb | 1.00 | 0.30 | 1.03 | 0.06 | 0.94 |
| 52 | Te |  |  | 2.20 | 0.03 |  |
| 53 | I  |  |  | 1.57 | 0.08 |  |
| 54 | Xe | 2.27 | 0.08 | -1.93 |  | 1.6E+04 |
| 55 | Cs |  |  | 1.10 | 0.02 |  |
| 56 | Ba | 2.17 | 0.07 | 2.20 | 0.03 | *0.94* |
| 57 | La | 1.14 | 0.03 | 1.19 | 0.02 | *0.88* |
| 58 | Ce | 1.61 | 0.06 | 1.60 | 0.02 | *1.02* |
| 59 | Pr | 0.76 | 0.04 | 0.78 | 0.03 | *0.96* |
| 60 | Nd | 1.45 | 0.05 | 1.47 | 0.02 | *0.96* |
| 62 | Sm | 1.00 | 0.05 | 0.96 | 0.02 | *1.10* |
| 63 | Eu | 0.52 | 0.04 | 0.53 | 0.02 | *0.97* |
| 64 | Gd | 1.11 | 0.05 | 1.07 | 0.02 | *1.10* |
| 65 | Tb | 0.28 | 0.10 | 0.34 | 0.03 | 0.88 |
| 66 | Dy | 1.13 | 0.06 | 1.15 | 0.02 | *0.96* |
| 67 | Ho | 0.51 | 0.10 | 0.49 | 0.03 | 1.04 |
| 68 | Er | 0.96 | 0.06 | 0.94 | 0.02 | *1.04* |
| 69 | Tm | 0.14 | 0.04 | 0.14 | 0.03 | *1.00* |
| 70 | Yb | 0.86 | 0.10 | 0.94 | 0.02 | 0.83 |
| 71 | Lu | 0.12 | 0.08 | 0.11 | 0.02 | *1.02* |
| 72 | Hf | 0.88 | 0.08 | 0.73 | 0.02 | *1.43* |
| 73 | Ta |  |  | -0.14 | 0.04 |  |
| 74 | W  | 1.11 | 0.15 | 0.67 | 0.04 | 2.75 |
| 75 | Re |  |  | 0.28 | 0.04 |  |
| 76 | Os | 1.45 | 0.11 | 1.37 | 0.03 | 1.21 |
| 77 | Ir | 1.38 | 0.05 | 1.34 | 0.02 | *1.10* |
| 78 | Pt | 1.74 | 0.30 | 1.64 | 0.03 |  |
| 79 | Au | 1.01 | 0.18 | 0.82 | 0.04 | 1.54 |
| 80 | Hg |  |  | 1.19 | 0.08 |  |
| 81 | Tl | 0.95 | 0.20 | 0.79 | 0.03 | 1.43 |
| 82 | Pb | 2.00 | 0.06 | 2.06 | 0.03 | *0.88* |
| 83 | Bi |  |  | 0.67 | 0.04 |  |
| 90 | Th | < 0.08 | UL | 0.08 | 0.03 | 1.01 |
| 92 | U  | < -0.47 | UL | -0.52 | 0.03 | 1.11 |

σ – standard deviation in dex: $(10^{(\sigma\ in\ dex)} - 1) \times 100 = \sigma$ in %;
meteorite data: log $N_{El}$+ 1.533 where $N_{El}$ = abundance relative to $10^6$ Si atoms (Table 1);
UL – upper limit;
Last column: Values in italics indicate the 39 elements used for calculating conversion factor.

**Tl**: The Tl value of 0.95±0.2 is from the linear average of the endmember composition of $0.72 \leq A(Tl) \leq 1.1$ selected in [03L] that was found for sunspot spectra. The uncertainty quoted here is to indicate the derived range. There are no new measurements.

**Th:** The photospheric Th abundance is difficult to determine because the only accessible and weak Th line in the photospheric spectrum is heavily blended with Ni I and Ni II. Caffau et al. [08C2] report a nominal abundance A(Th) = 0.08±0.03, which should not be over-interpreted because of the line blends.

### 3.4.5 Comparison of photospheric and meteoritic abundances

In order to compare the CI-chondrite abundances with the photospheric abundances, the Si-normalized atomic meteorite abundances of Table 3 ($N_{Si} = 10^6$; cosmochemical abundance scale) were converted to the hydrogen-normalized abundances (log $N_H$ = A(H) = 12; astronomical abundance scale) of Table 4. The conversion between the two scales requires a scale conversion factor (for the log-scales, a conversion constant). This factor was calculated by subtracting the logarithm of the Si-normalized meteoritic abundances (Table 3) from the logarithmic H-normalized solar abundances (Table 4). The comparison was made for all elements heavier than neon that have uncertainties of less than 0.1 dex, i.e., below at about 25 %, in their photospheric abundance determinations. There are 39 elements that qualify for this procedure. The log of the average ratio of *solar abundance per $10^{12}$ H atoms/meteorite abundance per $10^6$ Si atoms* is 1.533 ± 0.042. The cosmochemical and astronomical scales are coupled as:

$$A(X) = \log N_{astron}(X) = \log N_{cos}(X) + 1.533 \tag{1}$$

This yields a Si-abundance on the astronomical scale of A(Si) = 7.533 and a hydrogen abundance on the meteoritic scale of log $N_{met}$(H) = 10.47, corresponding to $2.93 \times 10^{10}$ atoms of H relative to $10^6$ atoms Si, i.e., there are about 30 000 more H atoms than Si atoms in the sun. A conversion constant of 1.554 ± 0.020 was used by Anders and Grevesse [89A], which leads to a H abundance of $2.79 \times 10^{10}$ atoms on the meteoritic scale. Lodders [03L] reported a value of 1.539 ± 0.046 for 35 elements for which photospheric abundances were determined with less than 25% uncertainty, and used a constant of 1.540, which is exactly the log of the ratio of Si in the astronomic to the meteoritic scale. The lower conversion factor found here is the result of a systematic decrease of the values reported for photospheric abundance since Anders and Grevesse [89A] made their compilation. In calculating their scale conversion factor, [89A] chose 12 elements that led to an uncertainty of only 5 % in the conversion factor, compared to the uncertainty of about 10% from the larger sample of elements used here and in [03L] to calculate the factor. It is also important that the elements used in the factor calculation span a larger range in different elemental properties (e.g., atomic number, mass, first ionization potential, condensation temperature) because the premise in linking the meteoritic and solar data is that there are no chemical and physical fractionations of the elements (except for the obvious loss of ultra volatile elements from meteorites). The small spread in the conversion factor nevertheless indicates that there is basic agreement of solar and meteoritic abundances. There is no apparent dependence of the conversion factor on atomic number, mass or any other nuclear or elemental property. We therefore believe that a reasonable estimate for the uncertainty of the relative scale of solar and meteoritic abundances is about 10%.

Table 4 compares the solar and CI-chondrite abundances given on the astronomical abundance scale; equation (1) was applied to the meteoritic abundances of Table 3. The last column in Table 4 shows photosphere/CI-chondrite abundance ratios. A more systematic comparison of the photospheric and meteoritic data is done in Table 5, where elements are arranged in order of the absolute differences between Sun and Orgueil on the logarithmic scale, so elements with the largest relative abundance differences are listed first. The group of elements strongly depleted in Orgueil relative to the Sun consists of elements occurring predominantly as gases under planetary conditions. The largest depletion factors are found for the rare gases which decrease with increasing atomic number. The depletion sequence for N, C, and H indicates the general rarity of solid nitrogen compounds in meteorites and the predominance of oxides and silicates.

Only Li is clearly consumed in the interior of Sun by nuclear reactions. Nominally, B is depleted in the Sun by about 20 % but within the stated uncertainties, it may not be affected. Beryllium is another fragile element like Li and B, and may be subject to destruction in the Sun. However, the comparison of photospheric and CI-chondritic abundances indicates that there was no Be loss in the Sun; indeed; the nominal Be abundance for the Sun is higher than in CI-chondrites.

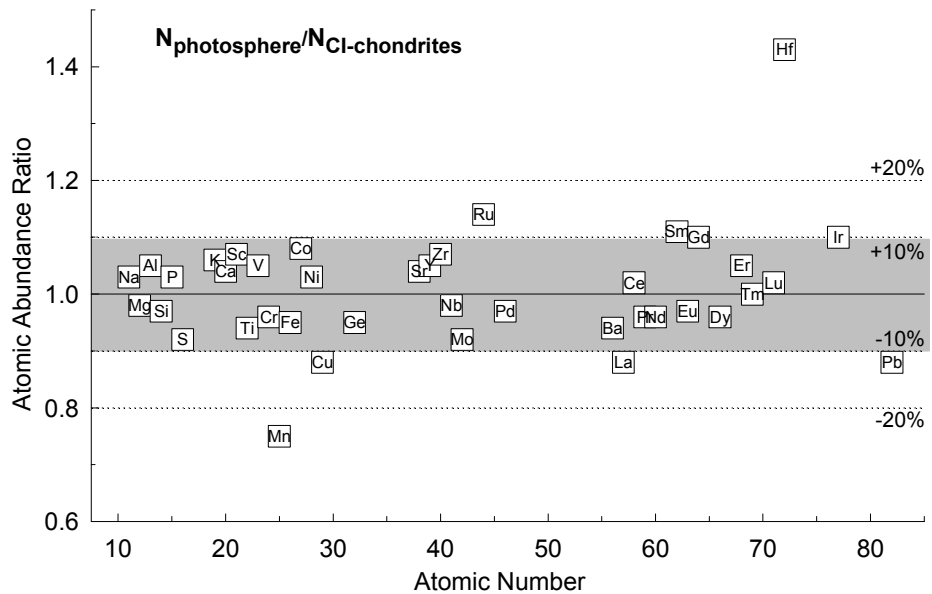

**Fig. 4.** Comparison of photospheric and CI-chondrite abundances, excluding highly volatile elements. Only elements with uncertainties below ~ 25% (<0.10 dex) in the photospheric abundance determinations are shown. The grey-shaded area is for elements that agree within ±10%; the range for 20% agreement is shown by the other dotted lines. Hafnium and Mn are well determined in meteorites and in the Sun, but differ by more than 30 %. Lithium and Be would also qualify for inclusion in this diagram but are not included. Lithium is by a factor of more than 100 higher in meteorites, because it is destroyed in the interior of the Sun by nuclear reactions. Beryllium only differs by 15%, suggesting less or no destruction by nuclear reactions.

There are 21 other elements where the difference between photospheric and Orgueil abundances exceeds 10%. For most of these elements, the combined uncertainty of the photospheric and meteoritic determinations is larger than the difference in abundance, indicating that solar and meteoritic abundances agree within error limits. Problematic elements are W, Rb, Ga, Hf, and Mn, which have abundance differences larger than the combined error bars. The error bar assigned to the solar abundance determinations of W is 40%, and for Rb and Ga 25%, which does not cover the abundance differences, which are a factor of 2.76 for W, 1.64 for Rb, and 0.61 for Ga. The estimated uncertainties for the concentrations of these three elements in CI-chondrites are below 10%. We suspect either that the uncertainties of the photospheric determinations are underestimated, overestimated or that there are unknown systematic errors involved. The abundances of Ga, Rb, and W need to be re-determined in the photosphere. The case is more serious for the two elements Hf and Mn. The abundance differences are 40 and 30%, but the assigned errors are 20 % for Hf and 12% for Mn. Manganese was recently re-determined in the Sun by several groups, however, it remains a factor of 1.35 times higher in Orgueil than in the Sun. Line blending may not be the culprit as this usually leads to over-estimated abundances for the photosphere (e.g., like for Indium); but the photospheric value is lower than in CI-chondrites. Manganese can be accurately measured in meteorites. Its concentration in Orgueil is similar to that in two other CI-chondrites, Alais and Ivuna [03L], and it fits with the abundances of other elements of similar volatility. The Mn/Na ratio is, for example, constant in most carbonaceous chondrites [03P2], despite decreasing abundances of both elements with increasing petrologic type (from CI to CV) and despite the very different geochemical behavior of Mn and Na. We suspect that an unidentified problem in the photospheric abundance analysis may cause the discrepancy of the meteoritic and solar Mn abundances.

The problem is reversed for Hf: the meteoritic abundance of Hf is less than the photospheric one. This could indicate a problem with the photospheric abundance determination and suggest that line blending is more severe than already corrected for in current models. Two recent Hf analyses using different models essentially obtain the same abundance, and if there is a problem with the analysis, it remains elusive. The Hf concentration in CI-chondrites has been accurately determined, because Hf is important for Lu-Hf and W-Hf dating. The very constant Lu/Hf ratio discussed above closely ties Hf to other refractory elements which do not show such large differences in abundance to the sun as does Hf. This issue awaits resolution.

There are now 39 elements with a photospheric/CI-chondrite abundance ratio between 0.9 and 1.1 (Table 5 and Figs. 4 and 5), whereas in 2003, there were only 27 elements [03P1]. In particular, there is excellent agreement between the solar and meteoritic abundances of the REE, due to the recent improvements in photospheric measurements [09S2]. The agreement of the solar data with our new compilation is better than 5%, except for La and Tb, which deviate by 12 and 15%, respectively.

Special cases are Cr and S, where the combined error bars just match the abundance difference. The solar abundances of Cr and S have very low formal uncertainties of ± 2%, which are identical or below the estimated uncertainty of the meteoritic values, and the uncertainty estimates may be too optimistic. The abundance difference is below the combined error bars for all other elements in this group, as indicated by the negative sign in the last column of Table 5.

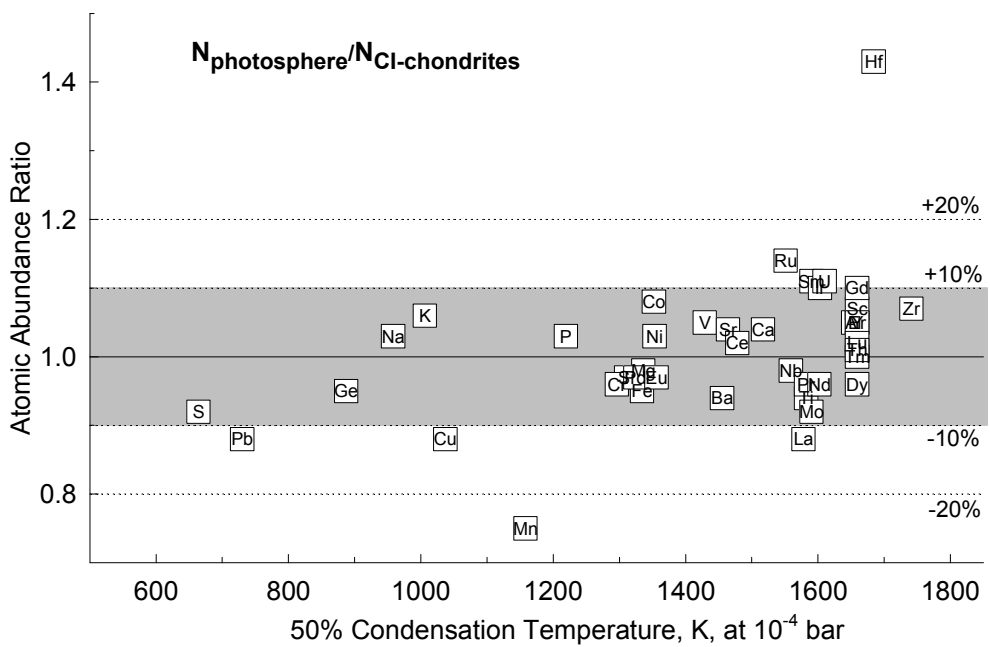

**Fig. 5.** Same as Fig. 4, but abundance ratios are plotted as function of 50% condensation temperatures. The agreement between solar and meteoritic abundances is independent of the volatility of the elements.

In summary, the agreement between photospheric and meteoritic abundances is further improved with the new solar and meteoritic data. As shown in Figs. 5 and 6 there is no apparent trend of the solar/meteoritic abundance ratios with increasing atomic number, or increasing volatility (decreasing condensation temperature) or any other property of nuclides or elements. In addition, the solar/meteoritic abundance ratios are independent of the geochemical character of an element, whether it is lithophile, siderophile or chalcophile, which indicates that any chemical and physical fractionation of silicates, metal, and sulfides did not affect CI-chondrite abundances.

**Table 5.** Comparison of elemental abundances in the Sun and the CI-chondrite Orgueil
(in order of decreasing difference in abundance)

|   | (1) $\log N_{EL}(Sun) - \log_{EL}(Orgueil)$ (absolute dex) | (2) $N_{EL}(Sun)/N_{EL}(Orgueil)$ (linear) | (3) combined uncertainty Sun + CI-chondrite (dex) | (4) difference (1)-(3) (dex) |
|---|---|---|---|---|
| **incompletely condensed elements in meteorites (rare gases, H, C, N, O)** | | | | |
| He | 9.61 | 4.1E+09 | 0.02 | 9.59 |
| Ne | 9.15 | 1.4E+09 | 0.10 | 9.05 |
| Ar | 6.98 | 9.6E+06 | 0.10 | 6.88 |
| Kr | 5.53 | 3.4E+05 | 0.08 | 5.45 |
| Xe | 4.20 | 1.6E+04 | 0.08 | 4.12 |
| H  | 3.76 | 5.7E+03 | 0.04 | 3.72 |
| N  | 1.58 | 38.4 | 0.18 | 1.40 |
| C  | 0.98 | 9.46 | 0.08 | 0.89 |
| O  | 0.31 | 2.06 | 0.11 | 0.20 |
| **elements consumed by nuclear reactions in the Sun** | | | | |
| Li | 2.18 | 0.007 | 0.15 | 2.03 |
| **elements with abundance differences between Sun and meteorites exceeding 10%** | | | | |
| *W*  | *0.44* | *2.76* | *0.19* | *0.25* |
| Ag | 0.28 | 0.52 | 0.32 | -0.04 |
| Cl | 0.25 | 1.79 | 0.36 | -0.11 |
| *Rb* | *0.22* | *1.64* | *0.13* | *0.09* |
| *Ga* | *0.22* | *0.61* | *0.12* | *0.09* |
| Au | 0.19 | 1.54 | 0.22 | -0.03 |
| Tl | 0.16 | 1.43 | 0.23 | -0.08 |
| *Hf* | *0.15* | *1.43* | *0.10* | *0.05* |
| *Mn* | *0.13* | *0.75* | *0.06* | *0.06* |
| F  | 0.12 | 1.32 | 0.36 | -0.24 |
| B  | 0.11 | 0.78 | 0.21 | -0.10 |
| Pt | 0.10 | 1.26 | 0.33 | -0.23 |
| Sn | 0.09 | 0.81 | 0.36 | -0.27 |
| Os | 0.08 | 1.21 | 0.14 | -0.06 |
| Yb | 0.08 | 0.83 | 0.12 | -0.04 |
| Be | 0.06 | 1.15 | 0.12 | -0.06 |
| Ru | 0.06 | 1.14 | 0.10 | -0.04 |
| Cu | 0.06 | 0.88 | 0.08 | -0.03 |
| Pb | 0.06 | 0.88 | 0.09 | -0.03 |
| Tb | 0.06 | 0.88 | 0.13 | -0.07 |
| *La* | *0.05* | *0.88* | *0.05* | *0.00* |

**Table 5.** Comparison of elemental abundances in the Sun and the CI-chondrite Orgueil
(in order of decreasing difference in abundance)

| | (1) log $N_{EL}$(Sun) – log $_{EL}$(Orgueil) (absolute dex) | (2) $N_{EL}$(Sun)/ $N_{EL}$(Orgueil) (linear) | (3) combined uncertainty Sun + CI-chondrite (dex) | (4) difference (1)-(3) (dex) |
|---|---|---|---|---|
| **elements with abundance difference between Sun and meteorites of 10% or below** | | | | |
| Sm | 0.04 | 1.10 | 0.07 | -0.03 |
| Gd | 0.04 | 1.10 | 0.07 | -0.03 |
| Ir | 0.04 | 1.10 | 0.07 | -0.03 |
| Cd | 0.04 | 1.10 | 0.14 | -0.10 |
| Rh | 0.04 | 1.09 | 0.16 | -0.12 |
| Mo | 0.04 | 0.92 | 0.09 | -0.05 |
| Co | 0.03 | 1.08 | 0.09 | -0.06 |
| Sc | 0.03 | 1.07 | 0.12 | -0.09 |
| Zr | 0.03 | 1.07 | 0.06 | -0.03 |
| Sb | 0.03 | 0.94 | 0.36 | -0.33 |
| Zn | 0.03 | 0.94 | 0.19 | -0.17 |
| Ba | 0.03 | 0.94 | 0.10 | -0.07 |
| Ti | 0.03 | 0.94 | 0.09 | -0.07 |
| *S* | *0.03* | *0.92* | *0.03* | *0.00* |
| K | 0.02 | 1.06 | 0.05 | -0.03 |
| Ge | 0.02 | 0.95 | 0.09 | -0.07 |
| Y | 0.02 | 1.05 | 0.06 | -0.04 |
| Fe | 0.02 | 0.95 | 0.09 | -0.07 |
| V | 0.02 | 1.05 | 0.04 | -0.02 |
| Al | 0.02 | 1.05 | 0.08 | -0.06 |
| *Cr* | *0.02* | *0.96* | *0.02* | *0.00* |
| Er | 0.02 | 1.05 | 0.08 | -0.06 |
| Pr | 0.02 | 0.96 | 0.07 | -0.05 |
| Sr | 0.02 | 1.04 | 0.08 | -0.06 |
| Ho | 0.02 | 1.04 | 0.13 | -0.11 |
| Dy | 0.02 | 0.96 | 0.08 | -0.06 |
| Ca | 0.02 | 1.04 | 0.09 | -0.08 |
| Nd | 0.02 | 0.96 | 0.07 | -0.06 |
| Ni | 0.01 | 1.03 | 0.05 | -0.04 |
| Si | 0.01 | 0.97 | 0.07 | -0.06 |
| Eu | 0.01 | 0.97 | 0.06 | -0.05 |
| Pd | 0.01 | 0.97 | 0.06 | -0.05 |
| P | 0.01 | 1.03 | 0.08 | -0.07 |
| Na | 0.01 | 1.03 | 0.05 | -0.04 |
| Ce | 0.01 | 1.02 | 0.08 | -0.07 |
| Nb | 0.01 | 0.98 | 0.10 | -0.09 |
| Lu | 0.01 | 1.02 | 0.10 | -0.09 |
| Mg | 0.01 | 0.98 | 0.07 | -0.07 |
| Tm | 0.00 | 1.00 | 0.07 | -0.07 |

Note: (1) – sorted by decreasing difference between Sun and meteorites (absolute); (2) – atomic ratio Sun/meteorite; (3) – uncertainties of Sun and meteorite added; (4): Difference between solar and meteoritic abundances minus the combined error of the solar and meteoritic abundances.
Values in italics: If the difference is positive in col. (4), there are either problems with abundance data for the Sun or for meteorites, or the error assignments are too optimistic (see text for details).

### 3.4.6 Solar system abundances

#### 3.4.6.1 Recommended present-day solar abundances

We follow Lodders [03L] to derive the representative present-day solar abundance in Table 6 based on CI-chondrites, photospheric data, and theoretical calculations. In cases where solar and meteorite data have comparable accuracy for a given element, the recommended abundance is the average of these values. For other elements, meteoritic data seem more reliable. The recommended present day abundances in Table 6 are converted to the astronomical scale using the same conversion constant between the astronomical and cosmochemical abundance scales as described before. A graphical representation of the elemental abundance distribution is given in Fig. 6.

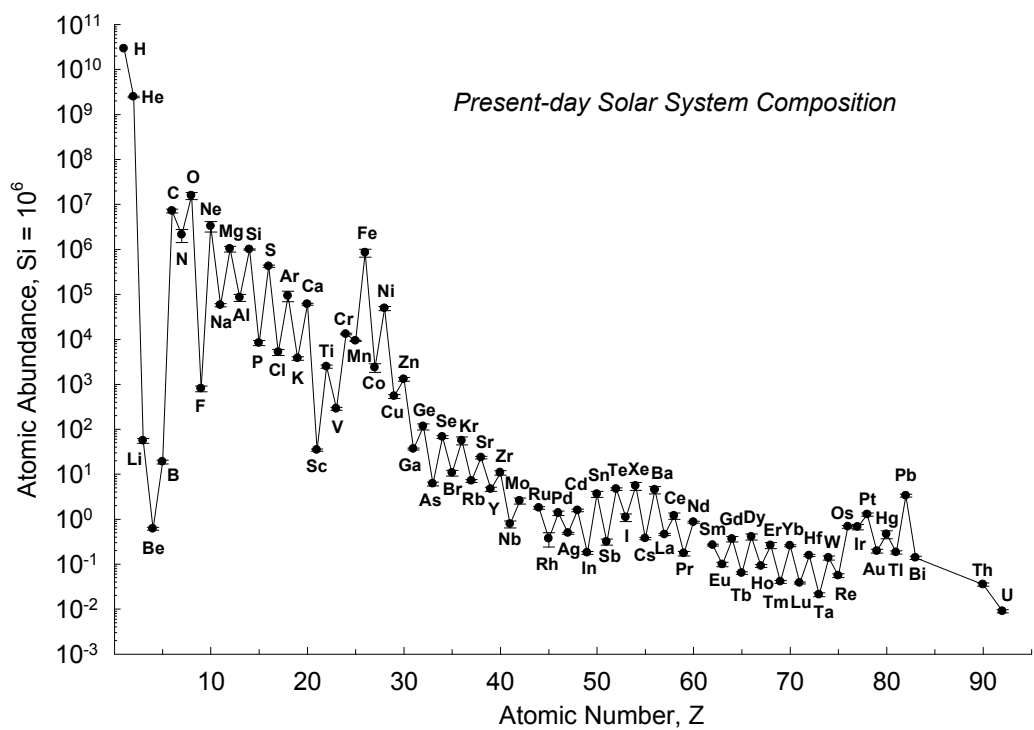

**Fig. 6.** Elemental present-day solar system abundances as function of atomic number normalized to $10^6$ Si atoms. The stability of $^{56}$Fe and the higher stability of elements with even atomic numbers are clearly visible.

The meteoritic abundances in Table 3 cannot be used for the extremely volatile elements, such as H, rare gases, C, N, and O which are strongly depleted in all types of meteorites. The recommended data for the abundances of ultra-volatile elements like H, C, N, and O are from photospheric determinations, and from various sources and theory for the noble gases (see comments for photospheric abundances of He, Ne and Ar above and for Kr and Xe in [03L]).

**Table 6.** Recommended present-day solar abundances from photospheric and meteoritic data

|    |    | source | N per $10^6$ Si atoms | σ | A (log N(H)=12) | σ |
|----|----|--------|------|------|------|------|
| 1  | H  | s      | 2.93E+10 |          | 12.00 |      |
| 2  | He | s,t    | 2.47E+09 | 0.12E+09 | 10.93 | 0.02 |
| 3  | Li | m      | 55.6     | 7.2      | 3.28  | 0.05 |
| 4  | Be | m      | 0.612    | 0.043    | 1.32  | 0.03 |
| 5  | B  | m      | 18.8     | 1.9      | 2.81  | 0.04 |
| 6  | C  | s      | 7.19E+06 | 0.69E+06 | 8.39  | 0.04 |
| 7  | N  | s      | 2.12E+06 | 0.68E+06 | 7.86  | 0.12 |
| 8  | O  | s      | 1.57E+07 | 0.28E+07 | 8.73  | 0.07 |
| 9  | F  | m      | 804      | 121      | 4.44  | 0.06 |
| 10 | Ne | s, t   | 3.29E+06 | 0.85E+06 | 8.05  | 0.10 |
| 11 | Na | a      | 57700    | 5100     | 6.29  | 0.04 |
| 12 | Mg | a      | 1.03E+06 | 0.15E+06 | 7.54  | 0.06 |
| 13 | Al | a      | 84600    | 15300    | 6.46  | 0.07 |
| 14 | Si | m      | 1.00E+06 | 0.03E+06 | 7.53  | 0.01 |
| 15 | P  | a      | 8300     | 1100     | 5.45  | 0.05 |
| 16 | S  | a      | 4.21E+05 | 0.24E+05 | 7.16  | 0.02 |
| 17 | Cl | m      | 5170     | 780      | 5.25  | 0.06 |
| 18 | Ar | s,t    | 92700    | 24000    | 6.50  | 0.10 |
| 19 | K  | a      | 3760     | 330      | 5.11  | 0.04 |
| 20 | Ca | m      | 60400    | 3000     | 6.31  | 0.02 |
| 21 | Sc | m      | 34.4     | 1.7      | 3.07  | 0.02 |
| 22 | Ti | m      | 2470     | 200      | 4.93  | 0.03 |
| 23 | V  | a      | 286      | 20       | 3.99  | 0.03 |
| 24 | Cr | a      | 13100    | 500      | 5.65  | 0.02 |
| 25 | Mn | m      | 9220     | 280      | 5.50  | 0.01 |
| 26 | Fe | a      | 8.48E+05 | 1.69E+05 | 7.46  | 0.08 |
| 27 | Co | a      | 2350     | 500      | 4.90  | 0.08 |
| 28 | Ni | a      | 49000    | 5000     | 6.22  | 0.04 |
| 29 | Cu | m      | 541      | 54       | 4.27  | 0.04 |
| 30 | Zn | m      | 1300     | 130      | 4.65  | 0.04 |
| 31 | Ga | m      | 36.6     | 1.8      | 3.10  | 0.02 |
| 32 | Ge | a      | 115      | 18       | 3.59  | 0.06 |
| 33 | As | m      | 6.10     | 0.55     | 2.32  | 0.04 |
| 34 | Se | m      | 67.5     | 4.7      | 3.36  | 0.03 |
| 35 | Br | m      | 10.7     | 1.6      | 2.56  | 0.06 |
| 36 | Kr | t      | 55.8     | 11.3     | 3.28  | 0.08 |
| 37 | Rb | m      | 7.10     | 0.50     | 2.38  | 0.03 |
| 38 | Sr | m      | 23.4     | 1.6      | 2.90  | 0.03 |
| 39 | Y  | a      | 4.63     | 0.50     | 2.20  | 0.04 |
| 40 | Zr | a      | 10.8     | 1.2      | 2.57  | 0.04 |
| 41 | Nb | a      | 0.780    | 0.139    | 1.42  | 0.07 |
| 42 | Mo | a      | 2.55     | 0.40     | 1.94  | 0.06 |
| 44 | Ru | m      | 1.78     | 0.11     | 1.78  | 0.03 |
| 45 | Rh | a      | 0.370    | 0.128    | 1.10  | 0.13 |
| 46 | Pd | a      | 1.36     | 0.15     | 1.67  | 0.04 |
| 47 | Ag | m      | 0.489    | 0.024    | 1.22  | 0.02 |
| 48 | Cd | m      | 1.57     | 0.11     | 1.73  | 0.03 |

**Table 6.** Recommended present-day solar abundances from photospheric and meteoritic data

|    |    | source | N per $10^6$ Si atoms | σ | A (log N(H)=12) | σ |
|----|----|--------|---------------------|-------|-----------------|------|
| 49 | In | m | 0.178 | 0.012 | 0.78 | 0.03 |
| 50 | Sn | m | 3.60 | 0.54 | 2.09 | 0.06 |
| 51 | Sb | m | 0.313 | 0.047 | 1.03 | 0.06 |
| 52 | Te | m | 4.69 | 0.33 | 2.20 | 0.03 |
| 53 | I  | m | 1.10 | 0.22 | 1.57 | 0.08 |
| 54 | Xe | t | 5.46 | 1.10 | 2.27 | 0.08 |
| 55 | Cs | m | 0.371 | 0.019 | 1.10 | 0.02 |
| 56 | Ba | a | 4.47 | 0.81 | 2.18 | 0.07 |
| 57 | La | m | 0.457 | 0.023 | 1.19 | 0.02 |
| 58 | Ce | a | 1.18 | 0.19 | 1.60 | 0.06 |
| 59 | Pr | a | 0.172 | 0.020 | 0.77 | 0.05 |
| 60 | Nd | m | 0.857 | 0.043 | 1.47 | 0.02 |
| 62 | Sm | m | 0.265 | 0.013 | 0.96 | 0.02 |
| 63 | Eu | a | 0.0984 | 0.0106 | 0.53 | 0.04 |
| 64 | Gd | a | 0.360 | 0.049 | 1.09 | 0.06 |
| 65 | Tb | m | 0.0634 | 0.0044 | 0.34 | 0.03 |
| 66 | Dy | a | 0.404 | 0.062 | 1.14 | 0.06 |
| 67 | Ho | m | 0.0910 | 0.0064 | 0.49 | 0.03 |
| 68 | Er | a | 0.262 | 0.042 | 0.95 | 0.06 |
| 69 | Tm | m | 0.0406 | 0.0028 | 0.14 | 0.03 |
| 70 | Yb | m | 0.256 | 0.013 | 0.94 | 0.02 |
| 71 | Lu | m | 0.0380 | 0.0019 | 0.11 | 0.02 |
| 72 | Hf | m | 0.156 | 0.008 | 0.73 | 0.02 |
| 73 | Ta | m | 0.0210 | 0.0021 | -0.14 | 0.04 |
| 74 | W  | m | 0.137 | 0.014 | 0.67 | 0.04 |
| 75 | Re | m | 0.0554 | 0.0055 | 0.28 | 0.04 |
| 76 | Os | m | 0.680 | 0.054 | 1.37 | 0.03 |
| 77 | Ir | a | 0.672 | 0.092 | 1.36 | 0.06 |
| 78 | Pt | m | 1.27 | 0.10 | 1.64 | 0.03 |
| 79 | Au | m | 0.195 | 0.019 | 0.82 | 0.04 |
| 80 | Hg | m | 0.458 | 0.092 | 1.19 | 0.08 |
| 81 | Tl | m | 0.182 | 0.015 | 0.79 | 0.03 |
| 82 | Pb | m | 3.33 | 0.23 | 2.06 | 0.03 |
| 83 | Bi | m | 0.138 | 0.012 | 0.67 | 0.04 |
| 90 | Th | m | 0.0351 | 0.0028 | 0.08 | 0.03 |
| 92 | U  | m | 0.00893 | 0.00071 | -0.52 | 0.03 |

Data sources: a = average of solar and meteoritic values; m = meteoritic based on CI-chondrites; s = photospheric or other solar data; t = theoretical considerations necessary.
Data on the astronomical scale (A(H) = log N(H)=12) and cosmochemical scale (N(Si)=$10^6$) are convertible through: A(X) = log N(X) + 1.533
Uncertainties U are convertible through: $(10^{\sigma \text{ (in dex)}} - 1) \times 100 = \sigma$ (in %)

**3.4.6.2 Mass fractions X, Y, and Z in present-day solar material**

Many applications in planetary sciences and astronomy require mass fractions of the elements rather than atomic abundances. The mass fraction of H is usually abbreviated as X, that of He as Y, and the sum of the mass fractions of all other heavy elements as Z. The overall sum of these mass fractions is X+Y+Z=1. Absolute mass fractions of X, Y, and Z can be derived if the ratio of Z/X is known from atomic abundance analysis (the Z/X ratio can always be computed without knowing the He abundance), and if either the mass fraction of H or He is known independently.

The mass fraction of He can be obtained from inversion of helioseismic data to match sound speeds of H and He dominated mixtures under physical conditions appropriate for the solar convection zone. The mass fraction Z, which combines all other heavy elements, is important as it governs opacities and thus the density structure of the Sun's outer convection zone. The depth of the solar convection zone derived from helioseismic data poses constraints on the permissible fraction of heavy elements (see [08B2] for a detailed review). The helioseismic inversion models require Z/X ratios and heavy element abundances (for opacities) as input. Therefore, the He mass fraction and the He abundance from such models are dependent of X/Z. In principle, one should use the abundance data and X/Z from the new compilation here to derive the He abundance from helioseismic models and fits to solar data, but modeling the solar interior is beyond the scope of this paper.

The absolute fractions of X, Y, and Z can also be derived from Z/X and the X (hydrogen mass fraction). Basu & Anita [04B2], [08B2] found that the estimated mass faction of H from helioseismic models is relatively independent on the Z/X ratios within the range of $0.0171 \leq Z/X \leq 0.0245$. If X is approximately constant for this range, variations in Z/X are solely due to variations in Z. In that case and because X+Y+Z = 1, the variations in Z must be accompanied by a corresponding variation in Y. The models by Basu & Anita [04B2] yield an average X=0.7389±0.0034. We assume here that compositional differences *within* Z (mainly governed by the mass fractions of O, C, Ne, see below) from our new abundance set also do not alter this model result by much, and use X from [04B2] to estimate the He mass fraction. With Z/X = 0.0191 found from the abundances in Table 6, and X = 0.739, one obtains Z = 0.0141, and from this Y = 1–X–Z = 0.2469. This mass fraction of He corresponds to an atomic He abundance of A(He) = 10.925, which is listed in Table 7.

The mass faction of heavy elements (Z=0.014) is intermediate to those in the compilations by [98G] (Z=0.017) and [05A1] (Z=0.012); see Table 7 for a comparison of present-day solar values. The He mass fraction (Y) found here is smaller than that in previous compilations, but the (rounded) He abundance is the same in the three compilations (Table 7). The H mass fraction of X=0.739 ([04B2] adopted here is essentially the same as in [05A1] and [07G], and the smaller value from [98G] appears to be the result of different model assumptions. The He abundance proposed here needs to be evaluated with results from helioseismic models calibrated with the recommended abundances of the elements (other than He) and the X/Z ratio found here.

**Table 7.** Present-day solar mass fractions and He abundance

| Present-Day:   | Z/X    | X      | Y      | Z      | A(He)  |
|----------------|--------|--------|--------|--------|--------|
| this work      | 0.0191 | 0.7390 | 0.2469 | 0.0141 | 10.925 |
| [05A1], [07G]  | 0.0165 | 0.7392 | 0.2486 | 0.0122 | 10.93  |
| [98G]          | 0.0231 | 0.7347 | 0.2483 | 0.0169 | 10.93  |

The Z/X and Z obtained here are higher than in the compilations by [03L] and Grevesse and co-workers, e.g., [05A1], [07G], but are still lower than those in the older compilation by [98G]. Table 7 compares the mass fractions of the most abundant elements in present-day solar material. About half of the mass fraction of Z comes from O, followed by C, Ne, and Fe (see Table 8). The higher Z here is mainly from the increased O and Ne abundances, which may help to eradicate the problem of the incompatibility of standard solar models with more recently recommended present-day solar abundances [08B2].

**Table 8.** Present-day solar composition (mass %)

|  | this work | [05A1,07G] | [98G] |
|---|---|---|---|
| H (=X) | 73.90 | 73.92 | 73.47 |
| He (=Y) | 24.69 | 24.86 | 24.83 |
| O | 0.63 | 0.54 | 0.79 |
| C | 0.22 | 0.22 | 0.29 |
| Ne | 0.17 | 0.10 | 0.18 |
| Fe | 0.12 | 0.12 | 0.13 |
| N | 0.07 | 0.06 | 0.08 |
| Si | 0.07 | 0.07 | 0.07 |
| Mg | 0.06 | 0.06 | 0.07 |
| S | 0.03 | 0.03 | 0.05 |
| all other elements | 0.04 | 0.02 | 0.04 |
| total heavy elements (=Z) | 1.41 | 1.22 | 1.69 |

### 3.4.6.3 Solar system abundances

The present-day photospheric abundances described above are different from those in the Sun 4.56 Ga ago, at the beginning of the solar system. Two processes affected the solar abundances over time. The first is element settling from the solar photosphere into the Sun's interior; the second is decay of radioactive isotopes that contribute to the overall atomic abundance of an element. The first is more important for large scale modeling; the second changes in isotopic compositions and their effects on abundances are comparably minor but important for radiometric dating. The isotopic effects are considered in the solar system abundance table in this section, and elemental abundance changes due to isotopic decay are minor.

Settling or diffusion of heavy elements from the photosphere to the interior boundary layer of the convection zone and beyond lowered the elemental abundances (relative to H) from protosolar values (see [08B2]). Diffusion reduced abundances of elements heavier than He by ~13% in the outer convection zone from original protosolar values, whereas the He abundance dropped by about ~15%; modeling these depletions also depends on opacities, hence on abundances. With these estimates, the proto-solar abundances (subscript 0) are calculated from the present-day data for the astronomical scales in logarithmic "dex" units as

$$A(He)_0 = A(He) + 0.061,$$

and for all elements heavier than He it is

$$A(X)_0 = A(X) + 0.053.$$

**Table 9.** Protosolar mass fractions and He abundance at the beginning of the solar system 4.56 Ga ago

|  | $Z_0/X_0$ | $X_0$ | $Y_0$ | $Z_0$ | $A(He)_0$ |
|---|---|---|---|---|---|
| this work [a] | 0.0215 | 0.7112 | 0.2735 | 0.0153 | 10.986 |
| [05A1], [07G] [b] | 0.0185 | 0.7133 | 0.2735 | 0.0132 | 10.985 |
| [98G] [c] | 0.0231 | 0.7086 | 0.2750 | 0.0163 | 10.99 |
| [89A] | 0.0267 | 0.7068 | 0.2743 | 0.0189 | 10.99 |

[a] Conversion of present-day to protosolar: $A(He)_0 = A(He) + 0.061$ dex, all other elements (except H) $A(X)_0 = A(X) + 0.053$ dex
[b] $A(He)_0 = A(He) + 0.057$ dex, all other elements (except H) $A(X)_0 = A(X) + 0.05$ dex.
[c] No diffusion effect on Z, $Z/X = Z_0/X_0$; (~10% loss of He).

On the cosmochemical abundance scale, the relative proto-solar abundances of the heavy elements appear unchanged from present-day solar values because that scale is normalized to Si. However, on this scale, the relative H abundance is less, and the He abundance is slightly higher (because of the higher diffusive loss). The protosolar mass fractions and He abundance at the time of solar system formations are listed in Table 9; which is to be compared to the present-day values Table 7. As with present day solar abundances, it is up to the solar models and helioseismology to derive the best-fitting current and proto solar He mass fractions from the given abundances of the other elements.

### 3.4.6.4 Abundances of the nuclides

The abundance of an element is determined by the numbers and abundances of its stable isotopes, which in turn depends on the stability and production rates of the nuclei during thermonuclear reactions in stellar interiors. In 1917, Harkins [17H] observed that elements with even atomic numbers are more abundant than their odd-numbered neighbors, which finds its explanation in the nuclear properties of the elements. There are 280 naturally occurring nuclides that make up the 83 stable and long-lived elements. These are all the elements up to Bi with Z =83, except for unstable Tc (Z=43) and Pm (Z=61) that only have short-lived isotopes, but the long-lived (relative to the age of the solar system) Th and U bring the total back to 83 elements.

The mass numbers of the stable and long-lived nuclides range from A=1 ($^1$H) to 209 ($^{209}$Bi) except for gaps at A=5 and 8. After $^{209}$Bi, we only have longer-lived nuclides of the actinides Th and U with the mass numbers 232, 235, and 238. Several nuclides have the same mass number but are isotopes of different elements, simply because A is given by Z+N. In comparisons of the nuclide distributions as function of mass number, the nuclides with the same A (isobaric nuclides, or isobars) are often summed up. Figure 7 shows the abundance distribution of the nuclides at the time of solar system formation 4.56 Ga ago. The abundances of nuclides with even A have usually higher abundances than the odd numbered nuclides that plot parallel to the even numbered A in a somewhat smoother distribution curve.

Abundances peak at mass numbers for closed proton and neutron "shells". These nuclear "shells" are to be understood analogous to the closed electron shells that characterize atomic properties. The "magic numbers" for nuclear stability are 2, 8, 20, 28, 50, 82, and 126; and nuclides with Z and/or N equal to these magic numbers are the ones that show large abundances in the diagram of abundance versus mass number (A=Z+N). This is particularly notable for the light doubly-magic nuclei with equal magic Z and N, e.g., $^4$He (Z=N=2), $^{16}$O(Z=N=8), and $^{40}$Ca (Z=N=20). Beyond the region of nuclides with mass numbers of 56 (the "Fe-peak" region), abundances decline more or less smoothly and spike at certain mass number regions. The nuclides beyond the Fe peak are products from neutron capture processes and the peaks correspond to regions where either nuclides are preferentially made by the slow-neutron capture, s-process operating in red giant stars (e.g., Y and Ba regions) or by the rapid-neutron capture r-process probably operating in supernovae (e.g., Pt region); see [97W], [02W], [08S3] for reviews on stellar nucleosynthesis. Here "slow" and "rapid" are in reference to beta-decay timescales of the intermediate, unstable nuclei produced during the neutron capture processes. The nuclide yields from these processes depend on the neutron energies and flux, but also on the abundance and stability of the target nuclei against neutron capture which in turn depends on Z and N. Hence, the abundance distribution becomes controlled by the more stable "magic" nuclides that serve as bottlenecks for the overall yields in the neutron capture processes.

Only a few nuclides beyond the Fe-group are exclusively produced by the s or r process; most of the nuclides have varying abundance contributions from both processes. If the fraction from each process to each isotope is known, the overall contribution of the r- and s-process to the elemental abundance can be estimated (see 08S3] for a recent table on the r- and s-contributions). A small number of proton-rich nuclides cannot be produced by neutron capture reactions and they are made by the p-process, originally thought to be a proton capture process. Currently discussed mechanisms of the p-process are photon induced reactions with high γ-fluxes that preferentially remove neutrons from target nuclei, or neutrino driven reactions, as well other nuclear processes. The p-process is not yet really understood. The contributions from p-process nuclides to elemental abundances are usually quite small, except for Mo, where p-process isotopes contribute about 25% to the abundance.

Table 10 is an update to the nuclide abundance table in [03L]. It gives the percent contribution of the isotope(s) for each element, and the atomic abundance relative to $10^6$ silicon atoms at the time of solar system formation. The abundances of radioactive isotopes (indicated by a star next to the element symbol) are adjusted accordingly from the measured present-day abundances. Table 10 contains several new measurements of isotopic compositions, including: Mo ([07W]), Dy ([02S1], [01C], Er ([98C]), Yb ([06D2]), and Lu ([06D3]).

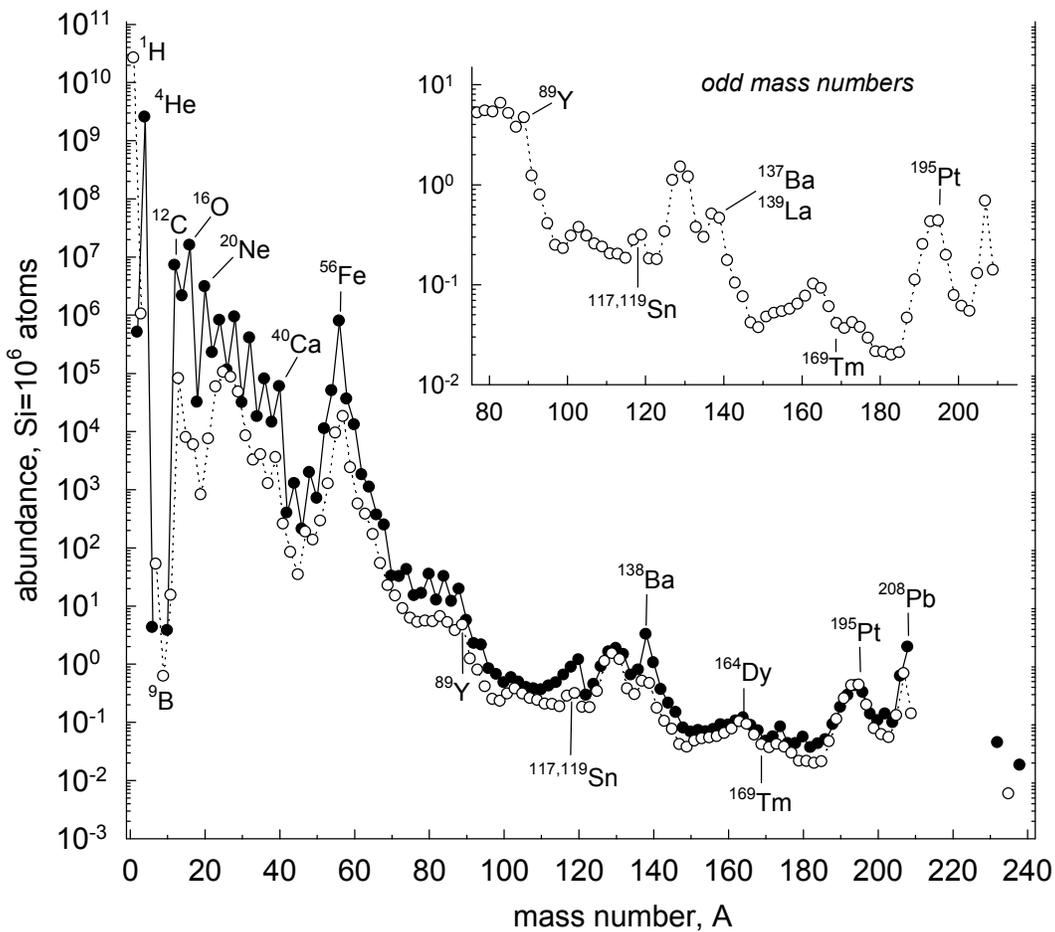

**Fig. 7.** Nuclide abundances plotted versus mass number. Open symbols are for odd mass numbers and full symbols are for even ones. The odd numbered nuclides form an approximately smooth curve (insert in Figure). Suess [47S] attached significance to this smoothness and suggested that elements with poorly known abundances may be found by interpolation. The abundance curve has, however, kinks (e.g. at Sn) inconsistent with smooth abundances.

**Table 10.** Nuclide abundances 4.56 Ga ago (normalized to $N(Si) = 10^6$ atoms)

| Z | | A | atom% | N | Z | | A | atom% | N |
|---|---|---|---|---|---|---|---|---|---|
| 1 | H | 1 | 99.9981 | 2.59E+10 | 20 | Ca | 46 | 0.004 | 2 |
| 1 | H | 2 | 0.00194 | 5.03E+05 | 20 | Ca | 48 | 0.187 | 113 |
| | | | 100 | 2.59E+10 | | | | 100 | 60400 |
| 2 | He | 3 | 0.0166 | 1.03E+06 | 21 | Sc | 45 | 100 | 34.4 |
| 2 | He | 4 | 99.9834 | 2.51E+09 | 22 | Ti | 46 | 8.249 | 204 |
| | | | 100 | 2.51E+09 | 22 | Ti | 47 | 7.437 | 184 |
| 3 | Li | 6 | 7.589 | 4.2 | 22 | Ti | 48 | 73.72 | 1820 |
| 3 | Li | 7 | 92.411 | 51.4 | 22 | Ti | 49 | 5.409 | 134 |
| | | | 100 | 55.6 | 22 | Ti | 50 | 5.185 | 128 |
| 4 | Be | 9 | 100 | 0.612 | | | | 100 | 2470 |
| 5 | B | 10 | 19.820 | 3.7 | 23 | V | 50 | 0.2497 | 0.7 |
| 5 | B | 11 | 80.180 | 15.1 | 23 | V | 51 | 99.7503 | 285.7 |
| | | | 100 | 18.8 | | | | 100 | 286.4 |
| 6 | C | 12 | 98.889 | 7.11E+06 | 24 | Cr | 50 | 4.3452 | 569 |
| 6 | C | 13 | 1.111 | 7.99E+04 | 24 | Cr | 52 | 83.7895 | 11000 |
| | | | 100 | 7.19E+06 | 24 | Cr | 53 | 9.5006 | 1240 |
| 7 | N | 14 | 99.634 | 2.12E+06 | 24 | Cr | 54 | 2.3647 | 309 |
| 7 | N | 15 | 0.366 | 7.78E+03 | | | | 100 | 13100 |
| | | | 100 | 2.12E+06 | 25 | Mn | 55 | 100 | 9220 |
| 8 | O | 16 | 99.763 | 1.57E+07 | 26 | Fe | 54 | 5.845 | 49600 |
| 8 | O | 17 | 0.037 | 5.90E+03 | 26 | Fe | 56 | 91.754 | 7.78E+05 |
| 8 | O | 18 | 0.200 | 3.15E+04 | 26 | Fe | 57 | 2.1191 | 18000 |
| | | | 100 | 1.57E+07 | 26 | Fe | 58 | 0.2819 | 2390 |
| 9 | F | 19 | 100 | 804 | | | | 100 | 8.48E+05 |
| 10 | Ne | 20 | 92.9431 | 3.06E+06 | 27 | Co | 59 | 100 | 2350 |
| 10 | Ne | 21 | 0.2228 | 7.33E+03 | 28 | Ni | 58 | 68.0769 | 33400 |
| 10 | Ne | 22 | 6.8341 | 2.25E+05 | 28 | Ni | 60 | 26.2231 | 12900 |
| | | | 100 | 3.29E+06 | 28 | Ni | 61 | 1.1399 | 559 |
| 11 | Na | 23 | 100 | 57700 | 28 | Ni | 62 | 3.6345 | 1780 |
| 12 | Mg | 24 | 78.992 | 8.10E+05 | 28 | Ni | 64 | 0.9256 | 454 |
| 12 | Mg | 25 | 10.003 | 1.03E+05 | | | | 100 | 49000 |
| 12 | Mg | 26 | 11.005 | 1.13E+05 | 29 | Cu | 63 | 69.174 | 374 |
| | | | 100 | 1.03E+06 | 29 | Cu | 65 | 30.826 | 167 |
| 13 | Al | 27 | 100 | 8.46E+04 | | | | 100 | 541 |
| 14 | Si | 28 | 92.230 | 9.22E+05 | 30 | Zn | 64 | 48.63 | 630 |
| 14 | Si | 29 | 4.683 | 4.68E+04 | 30 | Zn | 66 | 27.9 | 362 |
| 14 | Si | 30 | 3.087 | 3.09E+04 | 30 | Zn | 67 | 4.1 | 53 |
| | | | 100 | 1.00E+06 | 30 | Zn | 68 | 18.75 | 243 |
| 15 | P | 31 | 100 | 8300 | 30 | Zn | 70 | 0.62 | 8 |
| 16 | S | 32 | 95.018 | 400258 | | | | 100 | 1300 |
| 16 | S | 33 | 0.75 | 3160 | 31 | Ga | 69 | 60.108 | 22.0 |
| 16 | S | 34 | 4.215 | 17800 | 31 | Ga | 71 | 39.892 | 14.6 |
| 16 | S | 36 | 0.017 | 72 | | | | 100 | 36.6 |
| | | | 100 | 421245 | 32 | Ge | 70 | 21.234 | 24.3 |
| 17 | Cl | 35 | 75.771 | 3920 | 32 | Ge | 72 | 27.662 | 31.7 |
| 17 | Cl | 37 | 24.229 | 1250 | 32 | Ge | 73 | 7.717 | 8.8 |
| | | | 100 | 5170 | 32 | Ge | 74 | 35.943 | 41.2 |
| 18 | Ar | 36 | 84.595 | 78400 | 32 | Ge | 76 | 7.444 | 8.5 |
| 18 | Ar | 38 | 15.381 | 14300 | | | | 100 | 115 |
| 18 | Ar | 40 | 0.024 | 22 | 33 | As | 75 | 100 | 6.10 |
| | | | 100 | 92700 | 34 | Se | 74 | 0.89 | 0.60 |
| 19 | K | 39 | 93.132 | 3500 | 34 | Se | 76 | 9.37 | 6.32 |
| 19 | K* | 40 | 0.147 | 6 | 34 | Se | 77 | 7.64 | 5.15 |
| 19 | K | 41 | 6.721 | 253 | 34 | Se | 78 | 23.77 | 16.04 |
| | | | 100 | 3760 | 34 | Se | 80 | 49.61 | 33.48 |
| 20 | Ca | 40 | 96.941 | 58500 | 34 | Se | 74 | 0.89 | 0.60 |
| 20 | Ca | 42 | 0.647 | 391 | 34 | Se | 76 | 9.37 | 6.32 |
| 20 | Ca | 43 | 0.135 | 82 | 34 | Se | 77 | 7.64 | 5.15 |
| 20 | Ca | 44 | 2.086 | 1260 | 34 | Se | 78 | 23.77 | 16.04 |

**Table 10.** Nuclide abundances 4.56 Ga ago (normalized to N(Si) = $10^6$ atoms) – continued

| Z | | A | atom% | N | Z | | A | atom% | N |
|---|---|---|---|---|---|---|---|---|---|
| 34 | Se | 80 | 49.61 | 33.48 | 48 | Cd | 111 | 12.8 | 0.201 |
| 34 | Se | 82 | 8.73 | 5.89 | 48 | Cd | 112 | 24.13 | 0.380 |
| | | | 100 | 67.5 | 48 | Cd | 113 | 12.22 | 0.192 |
| 35 | Br | 79 | 50.686 | 5.43 | 48 | Cd | 114 | 28.73 | 0.452 |
| 35 | Br | 81 | 49.314 | 5.28 | 48 | Cd | 116 | 7.49 | 0.118 |
| | | | 100 | 10.7 | | | | 100 | 1.57 |
| 36 | Kr | 78 | 0.362 | 0.20 | 49 | In | 113 | 4.288 | 0.008 |
| 36 | Kr | 80 | 2.326 | 1.30 | 49 | In | 115 | 95.712 | 0.170 |
| 36 | Kr | 82 | 11.655 | 6.51 | | | | 100 | 0.178 |
| 36 | Kr | 83 | 11.546 | 6.45 | 50 | Sn | 112 | 0.971 | 0.035 |
| 36 | Kr | 84 | 56.903 | 31.78 | 50 | Sn | 114 | 0.659 | 0.024 |
| 36 | Kr | 86 | 17.208 | 9.61 | 50 | Sn | 115 | 0.339 | 0.012 |
| | | | 100 | 55.8 | 50 | Sn | 116 | 14.536 | 0.524 |
| 37 | Rb | 85 | 70.844 | 5.121 | 50 | Sn | 117 | 7.676 | 0.277 |
| 37 | Rb* | 87 | 29.156 | 2.108 | 50 | Sn | 118 | 24.223 | 0.873 |
| | | | 100 | 7.23 | 50 | Sn | 119 | 8.585 | 0.309 |
| 38 | Sr | 84 | 0.5580 | 0.13 | 50 | Sn | 120 | 32.593 | 1.175 |
| 38 | Sr | 86 | 9.8678 | 2.30 | 50 | Sn | 122 | 4.629 | 0.167 |
| 38 | Sr | 87 | 6.8961 | 1.60 | 50 | Sn | 124 | 5.789 | 0.209 |
| 38 | Sr | 88 | 82.6781 | 19.2 | | | | 100 | 3.60 |
| | | | 100 | 23.3 | 51 | Sb | 121 | 57.213 | 0.179 |
| 39 | Y | 89 | 100 | 4.63 | 51 | Sb | 123 | 42.787 | 0.134 |
| 40 | Zr | 90 | 51.452 | 5.546 | | | | 100 | 0.313 |
| 40 | Zr | 91 | 11.223 | 1.210 | 52 | Te | 120 | 0.096 | 0.005 |
| 40 | Zr | 92 | 17.146 | 1.848 | 52 | Te | 122 | 2.603 | 0.122 |
| 40 | Zr | 94 | 17.38 | 1.873 | 52 | Te | 123 | 0.908 | 0.043 |
| 40 | Zr | 96 | 2.799 | 0.302 | 52 | Te | 124 | 4.816 | 0.226 |
| | | | 100 | 10.78 | 52 | Te | 125 | 7.139 | 0.335 |
| 41 | Nb | 93 | 100 | 0.780 | 52 | Te | 126 | 18.952 | 0.889 |
| 42 | Mo | 92 | 14.525 | 0.370 | 52 | Te | 128 | 31.687 | 1.486 |
| 42 | Mo | 94 | 9.151 | 0.233 | 52 | Te | 130 | 33.799 | 1.585 |
| 42 | Mo | 95 | 15.838 | 0.404 | | | | 100 | 4.69 |
| 42 | Mo | 96 | 16.672 | 0.425 | 53 | I | 127 | 100 | 1.10 |
| 42 | Mo | 97 | 9.599 | 0.245 | 54 | Xe | 124 | 0.129 | 0.007 |
| 42 | Mo | 98 | 24.391 | 0.622 | 54 | Xe | 126 | 0.112 | 0.006 |
| 42 | Mo | 100 | 9.824 | 0.250 | 54 | Xe | 128 | 2.23 | 0.122 |
| | | | 100 | 2.55 | 54 | Xe | 129 | 27.46 | 1.499 |
| 44 | Ru | 96 | 5.542 | 0.099 | 54 | Xe | 130 | 4.38 | 0.239 |
| 44 | Ru | 98 | 1.869 | 0.033 | 54 | Xe | 131 | 21.80 | 1.190 |
| 44 | Ru | 99 | 12.758 | 0.227 | 54 | Xe | 132 | 26.36 | 1.438 |
| 44 | Ru | 100 | 12.599 | 0.224 | 54 | Xe | 134 | 9.66 | 0.527 |
| 44 | Ru | 101 | 17.060 | 0.304 | 54 | Xe | 136 | 7.87 | 0.429 |
| 44 | Ru | 102 | 31.552 | 0.562 | | | | 100 | 5.457 |
| 44 | Ru | 104 | 18.621 | 0.332 | 55 | Cs | 133 | 100 | 0.371 |
| | | | 100 | 1.78 | 56 | Ba | 130 | 0.106 | 0.005 |
| 45 | Rh | 103 | 100 | 0.370 | 56 | Ba | 132 | 0.101 | 0.005 |
| 46 | Pd | 102 | 1.02 | 0.0139 | 56 | Ba | 134 | 2.417 | 0.108 |
| 46 | Pd | 104 | 11.14 | 0.1513 | 56 | Ba | 135 | 6.592 | 0.295 |
| 46 | Pd | 105 | 22.33 | 0.3032 | 56 | Ba | 136 | 7.853 | 0.351 |
| 46 | Pd | 106 | 27.33 | 0.371 | 56 | Ba | 137 | 11.232 | 0.502 |
| 46 | Pd | 108 | 26.46 | 0.359 | 56 | Ba | 138 | 71.699 | 3.205 |
| 46 | Pd | 110 | 11.72 | 0.159 | | | | 100 | 4.471 |
| | | | 100 | 1.36 | 57 | La* | 138 | 0.091 | 0.000 |
| 47 | Ag | 107 | 51.839 | 0.254 | 57 | La | 139 | 99.909 | 0.457 |
| 47 | Ag | 109 | 48.161 | 0.236 | | | | 100 | 0.457 |
| | | | 100 | 0.489 | 58 | Ce | 136 | 0.186 | 0.002 |
| 48 | Cd | 106 | 1.25 | 0.020 | 58 | Ce | 138 | 0.250 | 0.003 |
| 48 | Cd | 108 | 0.89 | 0.014 | 58 | Ce | 140 | 88.450 | 1.043 |
| 48 | Cd | 110 | 12.49 | 0.197 | 58 | Ce | 142 | 11.114 | 0.131 |

**Table 10.** Nuclide abundances 4.56 Ga ago (normalized to $N(Si) = 10^6$ atoms) - continued

| Z | | A | atom% | N | Z | | A | atom% | N |
|---|---|---|---|---|---|---|---|---|---|
| 58 | Ce | | 100 | 1.180 | 72 | Hf | 176 | 5.206 | 0.0081 |
| 59 | Pr | 141 | 100 | 0.172 | 72 | Hf | 177 | 18.606 | 0.0290 |
| 60 | Nd | 142 | 27.044 | 0.231 | 72 | Hf | 178 | 27.297 | 0.0425 |
| 60 | Nd | 143 | 12.023 | 0.103 | 72 | Hf | 179 | 13.629 | 0.0212 |
| 60 | Nd | 144 | 23.729 | 0.203 | 72 | Hf | 180 | 35.100 | 0.0547 |
| 60 | Nd | 145 | 8.763 | 0.075 | | | | 100 | 0.156 |
| 60 | Nd | 146 | 17.130 | 0.147 | 73 | Ta | 180 | 0.0123 | 2.6E-06 |
| 60 | Nd | 148 | 5.716 | 0.049 | 73 | Ta | 181 | 99.9877 | 0.0210 |
| 60 | Nd | 150 | 5.596 | 0.048 | | | | 100 | 0.0210 |
| | | | 100 | 0.856 | 74 | W | 180 | 0.120 | 0.0002 |
| 62 | Sm | 144 | 3.073 | 0.008 | 74 | W | 182 | 26.499 | 0.0363 |
| 62 | Sm* | 147 | 14.993 | 0.041 | 74 | W | 183 | 14.314 | 0.0196 |
| 62 | Sm* | 148 | 11.241 | 0.030 | 74 | W | 184 | 30.642 | 0.0420 |
| 62 | Sm | 149 | 13.819 | 0.037 | 74 | W | 186 | 28.426 | 0.0390 |
| 62 | Sm | 150 | 7.380 | 0.020 | | | | 100 | 0.137 |
| 62 | Sm | 152 | 26.742 | 0.071 | 75 | Re | 185 | 35.662 | 0.0207 |
| 62 | Sm | 154 | 22.752 | 0.060 | 75 | Re* | 187 | 64.338 | 0.0374 |
| | | | 100 | 0.267 | | | | 100 | 0.0581 |
| 63 | Eu | 151 | 47.81 | 0.0471 | 76 | Os | 184 | 0.020 | 0.0001 |
| 63 | Eu | 153 | 52.19 | 0.0514 | 76 | Os | 186 | 1.598 | 0.0108 |
| | | | 100 | 0.0984 | 76 | Os | 187 | 1.271 | 0.0086 |
| 64 | Gd | 152 | 0.203 | 0.0007 | 76 | Os | 188 | 13.337 | 0.0904 |
| 64 | Gd | 154 | 2.181 | 0.0078 | 76 | Os | 189 | 16.261 | 0.110 |
| 64 | Gd | 155 | 14.800 | 0.0533 | 76 | Os | 190 | 26.444 | 0.179 |
| 64 | Gd | 156 | 20.466 | 0.0736 | 76 | Os | 192 | 41.070 | 0.278 |
| 64 | Gd | 157 | 15.652 | 0.0563 | | | | 100 | 0.678 |
| 64 | Gd | 158 | 24.835 | 0.0894 | 77 | Ir | 191 | 37.272 | 0.250 |
| 64 | Gd | 160 | 21.864 | 0.0787 | 77 | Ir | 193 | 62.728 | 0.421 |
| | | | 100 | 0.360 | | | | 100 | 0.672 |
| 65 | Tb | 159 | 100 | 0.0634 | 78 | Pt* | 190 | 0.014 | 0.0002 |
| 66 | Dy | 156 | 0.056 | 0.0002 | 78 | Pt | 192 | 0.783 | 0.010 |
| 66 | Dy | 158 | 0.095 | 0.0004 | 78 | Pt | 194 | 32.967 | 0.420 |
| 66 | Dy | 160 | 2.329 | 0.0094 | 78 | Pt | 195 | 33.832 | 0.431 |
| 66 | Dy | 161 | 18.889 | 0.0762 | 78 | Pt | 196 | 25.242 | 0.322 |
| 66 | Dy | 162 | 25.475 | 0.1028 | 78 | Pt | 198 | 7.163 | 0.091 |
| 66 | Dy | 163 | 24.896 | 0.1005 | | | | 100 | 1.27 |
| 66 | Dy | 164 | 28.260 | 0.1141 | 79 | Au | 197 | 100 | 0.195 |
| | | | 100 | 0.404 | 80 | Hg | 196 | 0.15 | 0.001 |
| 67 | Ho | 165 | 100 | 0.0910 | 80 | Hg | 198 | 9.97 | 0.046 |
| 68 | Er | 162 | 0.139 | 0.0004 | 80 | Hg | 199 | 16.87 | 0.077 |
| 68 | Er | 164 | 1.601 | 0.0042 | 80 | Hg | 200 | 23.10 | 0.106 |
| 68 | Er | 166 | 33.503 | 0.088 | 80 | Hg | 201 | 13.18 | 0.060 |
| 68 | Er | 167 | 22.869 | 0.060 | 80 | Hg | 202 | 29.86 | 0.137 |
| 68 | Er | 168 | 26.978 | 0.071 | 80 | Hg | 204 | 6.87 | 0.031 |
| 68 | Er | 170 | 14.910 | 0.039 | | | | 100 | 0.458 |
| | | | 100 | 0.262 | 81 | Tl | 203 | 29.524 | 0.054 |
| 69 | Tm | 169 | 100 | 0.0406 | 81 | Tl | 205 | 70.476 | 0.129 |
| 70 | Yb | 168 | 0.12 | 0.0003 | | | | 100 | 0.182 |
| 70 | Yb | 170 | 2.98 | 0.0076 | 82 | Pb | 204 | 1.997 | 0.066 |
| 70 | Yb | 171 | 14.09 | 0.0361 | 82 | Pb | 206 | 18.582 | 0.614 |
| 70 | Yb | 172 | 21.69 | 0.0556 | 82 | Pb | 207 | 20.563 | 0.680 |
| 70 | Yb | 173 | 16.10 | 0.0413 | 82 | Pb | 208 | 58.858 | 1.946 |
| 70 | Yb | 174 | 32.03 | 0.0821 | | | | 100 | 3.306 |
| 70 | Yb | 176 | 13.00 | 0.0333 | 83 | Bi | 209 | 100 | 0.1382 |
| | | | 100 | 0.256 | 90 | Th* | 232 | 100 | 0.0440 |
| 71 | Lu | 175 | 97.1795 | 0.0370 | 92 | U* | 234 | 0.002 | 4.9E-07 |
| 71 | Lu* | 176 | 2.8205 | 0.0011 | 92 | U* | 235 | 24.286 | 0.0058 |
| | | | 100 | 0.0380 | 92 | U* | 238 | 75.712 | 0.0180 |
| 72 | Hf | 174 | 0.162 | 0.0003 | | | | 100 | 0.0238 |

## 3.4.7 Elemental abundances in neighboring stars in the Milky Way

The sun is member of a big stellar system, the Milky Way, a typical (barred) spiral galaxy that contains about $10^{11}$ solar masses of baryonic matter (and much more dark matter). About 90% of the baryonic matter component is in stars and 10% in the thin interstellar matter (ISM) between stars. The ISM has a clumpy structure; in the densest clumps, the molecular clouds, stars are formed by gravitational collapse from locally unstable regions. These stars live for some period while burning their nuclear fuels from H over He to C, and for the most massive stars ($M>8M_{Sun}$) finally up to the Fe group. At the end of their lives, stars return between 50% and 90% of their initial mass back to the ISM, enriching the ISM with the products of their nucleosynthetic processes. From this enriched matter new stars are formed. This cycling of matter between stars and the ISM determines the abundances of the elements heavier than He, the so-called metals in astronomical notation, in the ISM and in the stars formed thereof.

Like all spiral galaxies, the Milky Way started to form about 13 Ga ago by the collapse of primordial matter that contained H, He, and tiny traces of Li and nothing else. The first generation of stars, thought to be very massive and, therefore, not surviving to the present, was essentially metal free. The fractional abundance of the metals in stars, the metallicity Z, then increased gradually over time by the cycle of star formation and mass-return from essentially zero to the present super-solar metallicity found in some recently formed stars. Low mass stars like the Sun require for their formation some minimum metallicity in the ISM; the lowest metallicity stars presently known have $Z \sim 10^{-5} Z_{Sun}$.

The turn-around time of matter between stars and ISM is presently ~2 Ga at the distance of 8 kpc of the Sun from the galactic centre. Closer to the centre it is shorter because of higher matter density and a higher star formation rate (stars have higher metallicity), farther out it is longer (stars have lower metallicity). For this reason, abundances can be compared only for stars from a narrow range of distances from the galactic centre, since only then the matter from which they formed had essentially the same previous history. Because of mutual gravitational scattering stars may migrate over some distance range during their lifetime (for the Sun this is estimated to be ±1.5 kpc [96W,02S2]), and stars on orbits with high eccentricity may be observed at a radius markedly different from the galactocentric distance of its birthplace. Hence, any sample of stars even from a narrow range of distances is polluted to some (presently not quantifiable) extent with stars born at different places. This has to be observed if solar abundances are to be compared to abundances of nearby stars.

Abundances of the ISM, from which the stars form, cannot be measured directly because in the ISM the refractory elements are condensed into dust (cf. [96S]). One possibility to determine initial element abundances of stars at the instant of their formation is to determine atmospheric abundances from some kind of `young' stars that have not changed the surface abundances since their formation from interstellar matter. Best suited for this purpose are main sequence F and G stars from the galactic neighborhood, which show the kinematics of thin disk stars. There is no mixing between the burning zone and surface layers during the main sequence evolution of such stars. Hotter stars pose problems for abundance determinations by non-thermal excitation of excited atomic levels. Spectra of cooler stars are crowded by molecular lines, which make abundance determinations difficult. For B and in particular O stars rapid rotation of such stars induce mixing between burning zones and stellar surface, thus photospheric abundances do not reflect the elemental abundances at the time of star formation.

A number of accurate abundances determinations of main sequence F and G stars have been performed and results for four cosmochemical important elements (C, O, Mg, Si) are shown in Fig. 8. They span a wide range of metallicities and cover essentially the whole evolution time of the thin disk. Fig. 8 demonstrates the enrichment of the ISM by the matter cycle between stars and ISM. The scatter of observed abundances is not significantly higher than the error of abundance determinations (< ±0.05 ... ±0.15 dex), thus abundance deviations between coeval stars are probably small [04P,06D4]. Solar System abundances as compared to main sequence F and G stars of the same metallicity from the solar neighborhood do not show conspicuous peculiarities.

Table 11 shows average element abundances for F- and G-type stars from the solar neighborhood [05B3, 06B] and for stars with a metallicity range close to solar. These stars should have formed at about the time as the Sun and are expected to show very similar abundances. Comparison with solar abundances, which are given in the second column, show that indeed abundances are identical within error limits of abundance determinations. Also shown are average abundances of stars with an age of no more than a few Ga, if stellar ages have been determined. For such stars, one can assume that they formed from interstellar material with essentially the same properties as the present day ISM of the galactic neighborhood.

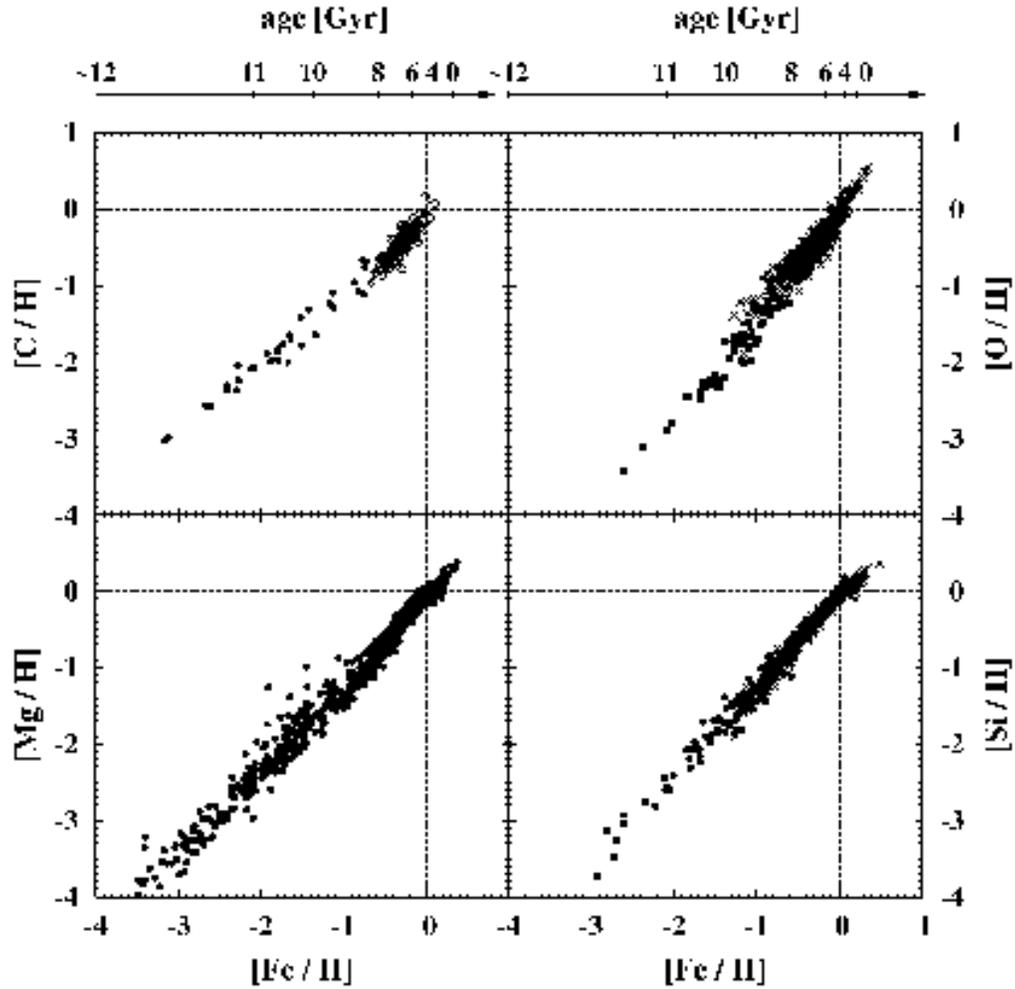

**Fig. 8.** Evolution of element abundances relative to H in the solar neighbourhood of some of the most abundant metals derived from main sequence G-stars. The range of metallicities -4 < [Fe/H] < 0.3 corresponds to stellar ages from ~12 Ga ([Fe/H] ~ -4) to less than 1 Ga ([Fe/H] ~0.2). The dashed lines correspond to solar abundances, i.e., abundances of the ISM 4.6 Ga ago. Solar abundances are within the range of observed abundances of other G-stars with same metallicity. Data sources: [04A2, 03R] for C, [00C, 05S2] for O, [04V] for Mg, [88G, 91G, 03G, 05S2] for Si.

Stars of early spectral type B have short lifetimes. Therefore, they sample abundances from the present day thin disk. Abundances have been determined in particular for B stars in stellar clusters and Table 11 shows average abundances for B dwarfs in stellar clusters with galactocentric distances from a range of ±2 kpc around the solar circle. Despite the rather heterogeneous observational material the scattering of observed abundances around the mean is moderate, i.e., element abundances in the ring 8±2 kpc around the galactic centre seem to be rather homogeneous. The average abundances and, thus, the metallicity Z, are typically slightly less than the present-day abundances found from main sequence F and G stars as it is found by Sofia & Meyer [01S].

Abundance determinations from HII regions around massive stars give also some information on the elemental abundances of the present day interstellar medium, but only a small number of elements can be analyzed. A serious problem is that the refractory elements are in part or even completely condensed into dust particles, but the degree of depletion usually cannot be determined with a sufficient degree of reliability, i.e., the total element abundance remains in most cases uncertain. Table 11 shows results from [04E] for the Orion nebula, since abundances in this nebula should reflect element abundances of the present day ISM in the solar vicinity. The Table includes already the estimated corrections with respect to dust depletion. The abundances are almost the same as for young F & G stars.

**Table 11.** Average abundances ($A$) of main sequence F and G stars with close to solar metallicity ($|\Delta[\text{Fe/H}]|<0.05$) and of young stars (age ~1 Ga) from the solar neighborhood, of main sequence B stars from a ±2 kpc wide ring around the solar circle, and abundances of the Orion nebula

| El. | Present-day solar system this work | Nearby F & G stars solar metallicity | | | Nearby F & G stars age < 1 Ga | | | B dwarfs | | | Orion nebula | | |
|---|---|---|---|---|---|---|---|---|---|---|---|---|---|
| | | $A$ | $\sigma_{abd}$ | $\sigma_*$ | $A$ | $\sigma_*$ | ref. | $A$ | $\sigma_*$ | ref. | $A$ | $\sigma_{abd}$ | ref. |
| He | 10.93 | | | | | | | 11.02 | 0.05 | 3 | 10.99 | 0.01 | 10 |
| C  | 8.39  | 8.37 | 0.06 | 0.11 | 8.39 | 0.11 | 2 | 8.32 | 0.10 | 4 | 8.52 | 0.05 | 10 |
| N  | 7.86  |      |      |      |      |      |   | 7.73 | 0.28 | 5 | 7.73 | 0.09 | 10 |
| O  | 8.73  | 8.75 |      | 0.07 | 8.77 | 0.13 | 1 | 8.63 | 0.18 | 4 | 8.73 | 0.09 | 10 |
| Ne | 8.05  |      |      |      |      |      |   | 7.97 | 0.07 | 6 | 8.05 | 0.07 | 10 |
| Na | 6.29  | 6.30 | 0.03 | 0.16 | 6.27 | 0.10 | 1 |      |      |   |      |      |    |
| Mg | 7.54  | 7.63 | 0.06 | 0.32 | 7.64 | 0.21 | 1 | 7.59 | 0.15 | 7 |      |      |    |
| Al | 6.46  | 6.52 | 0.05 | 0.24 | 6.54 | 0.22 | 1 | 6.24 | 0.14 | 5 |      |      |    |
| Si | 7.53  | 7.60 | 0.05 | 0.28 | 7.61 | 0.23 | 1 | 7.50 | 0.21 | 5 |      |      |    |
| S  | 7.16  | 7.17 | 0.16 |      | 7.29 | 0.10 |   | 7.22 | 0.10 | 4 | 7.22 | 0.04 | 10 |
| Ar | 6.50  |      |      |      |      |      |   | 6.63 | 0.06 | 8 | 6.62 | 0.05 | 10 |
| Ca | 6.31  | 6.42 | 0.03 | 0.37 | 6.48 | 0.39 | 1 |      |      |   |      |      |    |
| Ti | 4.93  | 4.92 | 0.03 | 0.11 | 4.94 | 0.14 | 1 |      |      |   |      |      |    |
| Cr | 5.65  | 5.66 | 0.02 | 0.13 | 5.73 | 0.28 | 1 |      |      |   |      |      |    |
| Fe | 7.46  | 7.55 | 0.06 | 0.12 | 7.61 | 0.25 | 1 | 7.46 | 0.08 | 9 |      |      |    |
| Ni | 6.22  | 6.22 | 0.02 | 0.09 | 6.25 | 0.07 | 1 |      |      |   |      |      |    |
| Zn | 4.65  | 4.53 | 0.06 | 0.27 | 4.54 | 0.13 | 1 |      |      |   |      |      |    |
| Z  | 0.0141 | 0.0141 | | | 0.0148 | | | 0.0119 | | | | | |

$\sigma_{abd}$ - accuracy of the individual abundance determinations from stellar spectra, $\sigma_*$ - scattering of the stellar abundances of the corresponding group of stars around the mean; $Z$ - metallicity estimated from these abundances.
Sun: Tables 6,7.
Data sources: (1)[05B3], (2)[06B], (3)[04L2], (4)[04D], (5)[00R], (6)[08M2], (7)[05L2], (8)[08L4], (9)[94C2], (10)[04E]